\documentclass[english,journal=jctcce,manuscript=article,email=true,etalmode=truncate,maxauthors=0,doi=true]{achemso}

\SectionNumbersOn

\usepackage{epstopdf}

\usepackage{graphicx}

\graphicspath{{figures/}}
\usepackage{physics}
\usepackage{multirow}
\usepackage{xcolor}
\usepackage{caption}
\usepackage{subcaption}
\usepackage{xcolor}
\usepackage{bm}
\usepackage{eufrak} 
\usepackage{yfonts} 
\usepackage[title,titletoc]{appendix}
\usepackage{diagbox}
\usepackage{multicol}
\usepackage{mhchem}
\usepackage{xspace}  

\usepackage{booktabs}

\usepackage{natmove}

\usepackage{hyperref} 
\hypersetup{colorlinks=true, citecolor=blue, urlcolor=blue, linkcolor=blue}
\usepackage{cleveref}
  	\crefname{figure}{Figure}{Figures}
  	\crefname{table}{Table}{Tables}
  	\crefname{equation}{Eq.}{Eqs.}
  	\crefname{section}{Section}{Sections}
  	\crefname{subsection}{Section}{Sections}
  	\crefname{subsubsection}{Section}{Sections}
  	\crefname{algorithm}{Algorithm}{Algorithms}

\usepackage[section]{minted}	

\newcommand\vartextvisiblespace[1][.5em]{%
  \makebox[#1]{%
    \kern.07em
    \vrule height.3ex
    \hrulefill
    \vrule height.3ex
    \kern.07em
  }
}
\usepackage{todonotes}

\def\ccsdptf{\ensuremath{\mathrm{CCSD}(2)_{\overline{\mathrm{F12}}}}\xspace}

\author{Samuel R. Powell}
\author{Edward F. Valeev}
\affiliation{Department of Chemistry, Virginia Tech, Blacksburg, Virginia, 24060}
\email{efv@vt.edu}

\title{Why are diffuse atomic orbitals needed for accurate electronic wave functions of even neutral molecules?} 

\begin{document}

\maketitle

\begin{abstract}
An accurate description of electron correlation energies in molecules requires either basis set extrapolation or the use of explicitly-correlated wave functions that address the deficiencies of standard determinantal expansions at short interelectronic distances. Practical and robust explicitly-correlated F12 methods require the use of standard or specialized atomic orbital (AO) basis sets that include diffuse AOs, even for neutral species. Although modern reduced-scaling formulations of explicitly-correlated many-body methods have become routinely applicable to molecules with hundreds of atoms, application of F12 methods to large molecular systems can be severely hampered due to the onset of ill-conditioning spurred by the presence of diffuse AOs in the F12-appropriate orbital basis sets. Here we re-examine why diffuse AOs are necessary for application of F12 methods. To help such an investigation, we developed a dual-basis formulation of traditional and F12 coupled-cluster singles and doubles (CCSD) methods in which the reference (occupied) and correlating (virtual) orbitals are expanded in separate AO basis sets. Our conclusion is that diffuse AOs are fundamentally important for the traditional (non-F12) description of dynamical correlation; the necessity of diffuse AOs in F12 calculations arises indirectly due to the dramatic reduction of the basis set error by the F12 terms such that the error due to the lack of diffuse AOs becomes comparable to the residual basis set incompleteness. The dual-basis CC methods are suggested as an important candidate formalism for  accurate (in particular, F12) reduced-scaling many-body methods in extended systems.
\end{abstract}

\section{Introduction}\label{sec:intro}
Basis set incompleteness error (BSIE) is often the dominant source of error in modern correlated models of electronic structure.\cite{VRG:tajti:2004:JCP,VRG:feller:2006:JCP}. The slow asymptotic basis set convergence of electronic energies, especially the correlation energy \cite{VRG:nyden:1981:JCP,VRG:kutzelnigg:1992:JCP,VRG:kutzelnigg:1992:JCPa}, is due to the inefficient modeling of the electron-electron cusp \cite{VRG:kato:1957:CPAM,VRG:pack:1966:JCP} by determinantal expansions. BSIE can be reduced by basis set extrapolation or inclusion of terms that depend on electronic distances explicitly (``explicit correlation''). Although basis set extrapolation is widely used,\cite{VRG:feller:1992:JCP,VRG:feller:1993:JCP,VRG:martin:1996:CPL,VRG:helgaker:1997:JCP,VRG:wilson:1997:JCP,VRG:halkier:1998:CPL,VRG:schwenke:2005:JCP} 
its empirical nature is a substantial weakness, and  
explicitly-correlated methods, such as the F12 methods \cite{VRG:kutzelnigg:1985:TCA,VRG:klopper:1987:CPL,VRG:noga:1997:RAiCC,VRG:ten-no:2004:CPL,VRG:valeev:2008:PCCP,VRG:valeev:2008:JCP,VRG:kong:2012:CR,VRG:ten-no:2012:TCA,VRG:ten-no:2012:WCMS} and transcorrelated methods\cite{VRG:yanai:2012:JCP,VRG:masteran:2023:JCP,VRG:ten-no:2023:JCP}, offer an attractive non-empirical alternative.

The F12 methods are currently the most practical explicitly-correlated formalism and have been applied to large molecules with hundreds of atoms and to infinite periodic crystals.\cite{VRG:pavosevic:2016:JCP,VRG:ma:2018:JCTC,VRG:kumar:2020:JCP,VRG:wang:2023:JCTC} Unfortunately, the application of F12 methods to large systems is hampered by the need to use basis sets specifically designed for use with F12 methods (cc-pV$X$Z-F12\cite{VRG:peterson:2008:JCP,VRG:peterson:2015:MP} and aug-cc-pV$X$Z-F12\cite{VRG:sylvetsky:2017:JCP}), which include a large number of diffuse atomic orbitals (AOs), or standard basis sets augmented by one or more sets of diffuse functions (e.g. aug-cc-pV$X$Z\cite{VRG:kendall:1992:JCP,VRG:woon:1993:JCP,VRG:woon:1994:JCP}), leading to poor numerical conditioning of the resulting equations, especially for large or periodic systems or for molecules in compressed geometries.\cite{VRG:maschio:2024:JPCA,VRG:motta:2017:PRX}
The preference for orbital basis sets containing diffuse AOs for F12 methods became apparent at the dawn of  ``modern'' F12 technology, when the use of the orbital basis for the many-electron integral RI was discontinued\cite{VRG:klopper:2002:JCP,VRG:valeev:2004:CPL,VRG:ten-no:2004:JCP} and short-range correlation factors were introduced\cite{VRG:ten-no:2004:CPL,VRG:may:2004:JCP,VRG:may:2005:PCCP,VRG:tew:2005:JCP}. Although Ten-no's pioneering work on the exponential F12 correlation factor\cite{VRG:ten-no:2004:CPL} did not specifically recommend the use of augmented orbital basis sets, significant error reduction due to their use was plain to see. Tew and Klopper were the first to explicitly recommend the use of augmented basis sets\cite{VRG:tew:2005:JCP} using their analysis of the MP2-F12 method (albeit based on the older orbital-invariant ansatz\cite{VRG:klopper:1991:CPL} rather than the SP ansatz\cite{VRG:ten-no:2004:JCP} whose use is now prevalent), noting that the optimal polarization orbitals are more diffuse than those included in Dunning's correlation consistent basis sets.\cite{VRG:dunning:1989:JCP,VRG:peterson:2008:JCP,VRG:peterson:2015:MP}
When  Peterson and coworkers designed the cc-pVXZ-F12 basis sets, which were tailored for F12 methods, they also noted the natural increase in the optimal spatial extent of AOs produced by optimization in presence of F12 terms. \cite{VRG:peterson:2008:JCP,VRG:peterson:2015:MP} It may even be necessary, especially for systems with noncovalent interactions, for excited states, and for response properties, to add further diffuse functions to these basis sets. \cite{VRG:sylvetsky:2017:JCP,VRG:yang:2009:JCP,VRG:yang:2009:JCPa,VRG:masteran:2025:JCTC}

It should not be entirely surprising that diffuse AOs are important for electron correlation in general and for F12 methods in particular. According to the known real-space asymptotics for Schr\"odinger eigenfunctions,\cite{VRG:ahlrichs:1981:PRA,VRG:carmona:1981:CP,VRG:carmona:1990:JFA} the rate of decay of the wave function into the vacuum depends on state-specific parameters, hence the wave functions must include parameters associated with diffuse AOs to be able to adjust the long-range decay. The F12 terms, which include only the \emph{universal} spin-dependent correlation physics associated with the electron-electron cusp,\cite{VRG:kato:1957:CPAM,VRG:pack:1966:JCP} will not be sufficient for the state-specific long-range behavior. The core design of the R12/F12 methods\cite{VRG:kutzelnigg:1985:TCA} reflects the idea that state-specific aspects of correlation require conventional orbital product expansions, with universal correlation physics modeled only by real-space correlators using a single fixed lengthscale parameter. This philosophy is in stark contrast with methods that use optimized real-space Jastrow factors which model not just short-range correlation effects (variational quantum Monte-Carlo\cite{VRG:foulkes:2001:RMP} and real-space transcorrelation methods\cite{VRG:boys:1969:PRSLA,VRG:boys:1969:PRSMPESa}). In particular, many recent applications of real-space transcorrelation employ Jastrow factors that include electron-electron (ee), electron-nucleus (en), and 3-particle (een) terms with many parameters\cite{VRG:schraivogel:2021:JCPa,VRG:kats:2024:FD} optimized in a state-specific manner (note that some argue against\cite{VRG:szenes:2024:FD} such complex multiphysics Jastrow factors). This allows such transcorrelated methods to use orbital basis sets without diffuse AOs, such as cc-pV$X$Z, with which they can outperform the parameter-free F12 methods, likely due to the need for diffuse AOs by the latter. Whether or not the use of such heavily-parametrized multiphysics  real-space Jastrow correlators can impact state universality and overall usability, these studies again stress the impetus to understand why diffuse AOs are needed for F12 methods. Such an understanding would also be key to the ongoing recent developments of transcorrelated F12 frameworks\cite{VRG:yanai:2012:JCP,VRG:ten-no:2023:JCP,VRG:masteran:2023:JCP,VRG:masteran:2025:JCTC,VRG:kumar:2020:JCP,VRG:kumar:2022:JCTC} that can be viewed as minimally parametrized transcorrelation methods.

Our recent observation that the optimal F12 geminals are substantially shorter in range when used with CCSD and higher-order correlation methods than in the context of MP2-F12\cite{VRG:powell:2025:JCTC} provided yet another reason to revisit the issue of diffuse AO in the F12 basis sets. However, the initial pragmatic motivation for this work was to determine whether the use of diffuse AOs can be avoided when computing the reference wave function for correlated F12 methods of large systems. To help answer these questions, we also developed a novel dual-basis (DB) coupled-cluster (CC) F12 method.

The key message of this work is that the need for diffuse AOs is not specifically related to the use of F12 terms; i.e., the impact of diffuse AOs on the correlation energy is only weakly affected by the presence of F12 terms. But it is the dramatic reduction of the basis set error resulting from the F12 terms that makes the relative importance of diffuse AOs much greater in the F12 methods than in their standard counterparts.
The use of the dual basis CC framework is also shown to be an effective way to introduce diffuse orbitals only to the correlated treatments, which may offer substantial savings in reduced-scaling F12 calculations of large systems.

The remainder of the paper is organized as follows. Section \ref{sec:formalism} describes the DB approach and the formalism that is used throughout. Section \ref{sec:technical} describes the technical details of the computational experiments that are reported and analyzed in \cref{sec:results}. \Cref{sec:summary} summarizes our findings and outlines future directions.
The appendices describe the notation (\cref{sec:notation}) and contain additional formalism details (\cref{sec:pvcabseqs}).

\section{Formalism}\label{sec:formalism}

The F12 methods rectify the lack of the electron-electron cusp in the determinantal expansions by the action of the {\em geminal} cluster operator:
\begin{align}
\hat{T}'_2 \equiv \frac{1}{4} \bar{R}^{ij}_{\alpha\beta} a^{\alpha\beta}_{ij}. 
\label{eq:t2prime}
\end{align}
with integrals
\begin{align}
  R^{ij}_{\alpha\beta} \equiv \langle \alpha\beta \vert (1-\hat{V}_1\hat{V}_2) f_{12} \vert ij \rangle,
  \label{eq:Rijab}
\end{align}
where $i,j$ are the (spin)orbitals occupied in the reference, and $\alpha,\beta$ their orthogonal complement in the formal complete basis (complete virtuals).
The integral kernel in \cref{eq:Rijab} involves the isotropic multiplicative correlation factor $f_{12}$; in this work $f_{12}$ is the Slater-type correlation factor:\cite{VRG:ten-no:2004:CPL}
\begin{align}
\label{eq:f12}
    f_{12} \equiv -\frac{\exp(-\gamma r_{12})}{\gamma}.
\end{align}

The diffuse AOs in the orbital basis set (OBS) enter the geminal cluster operator through (1) their contributions to the occupied orbitals (thus also changing the space of complete virtuals entering \cref{eq:t2prime}), and (2) their contributions to the OBS virtual (unoccupied) orbitals. The latter enter \cref{eq:Rijab} through $\hat{V} \equiv \sum_a \ket{a}\bra{a}$,
projector onto the OBS virtual orbitals, which through the projector $(1-\hat{V}_1\hat{V}_2)$ exclude the 2-body correlations accounted for by the conventional cluster operator (\cref{eq:T2}). Excluding diffuse AOs from the orbital basis will reduce the correlation energy by eliminating such orbitals from the conventional cluster operator, but this effect will be partially recovered by geminals by eliminating diffuse ``contributions'' to $\hat{V}$ in \cref{eq:Rijab}. The compensation will be partial, since the geminals are not designed to model the physics of ``diffuse'' 2-body correlations well. One of the goals of this work will be to quantify how well geminals can account for the diffuse 2-body correlation effects.

Let us now return to the effect of diffuse AOs on the occupied orbitals.
Although disagreements on various aspects of the basis set effects for mean-field models such as density functional theory (DFT) continue to this day,\cite{VRG:gray:2024:JPCA}
the consensus opinion seems to be that the inclusion of diffuse basis functions is important even for neutral species,\cite{VRG:jensen:2002:JCPa,VRG:lynch:2003:JPCA,VRG:papajak:2010:JCTC}
and especially for the properties that are sensitive to the quality of the ``tails'' of occupied orbitals as they decay into the vacuum, such as noncovalent interaction energetics\cite{VRG:witte:2016:JCP} and response properties\cite{VRG:jensen:2002:JCPa}.
Since such studies used limited varieties of basis sets and often did not distinguish basis set errors from model errors, some of their conclusions are at best not general. The recent availability of definitive numerical results for all-electron mean-field models with multiresolution real-space representations\cite{VRG:harrison:2004:JCP,VRG:yanai:2004:JCP,VRG:sekino:2008:JCP,VRG:harrison:2016:SJSC} allowed us to quantify the importance of diffuse AOs more precisely \cite{VRG:brakestad:2020:JCTC,VRG:sundahl:2021:,VRG:hurtado:2024:JCTC}. In particular, Ref. \citenum{VRG:sundahl:2021:} indicates that for absolute energies and atomization energies the diffuse AOs can have relatively minor effects, but for reaction energies the effects can be substantial.
Thus, to decompose the effects of diffuse AOs on the energies obtained with the F12 methods, it is desirable to be able to control their inclusion into the occupied orbitals (however minor such effects may be) independently of their inclusion into the correlating (virtual) basis.

To this end, we developed a {\em dual-basis} formulation of the conventional and explicitly-correlated coupled-cluster methods.
The dual-basis moniker refers to the fact that the reference (occupied) and the correlating (virtual/unoccupied) orbitals are expanded in two separate sets of AO basis functions. The only requirement is that the bases of occupied orbitals (OBS) and virtual orbitals (VBS) must be orthonormal and mutually orthogonal.

Havriliak and King pioneered the idea of dual basis methods by perturbatively compensating for the basis set error of the Hartree-Fock orbitals;\cite{VRG:havriliak:1983:JACS,VRG:havriliak:1984:CJP} this allowed for economical modeling of Rydberg states without the need to incorporate the crucial diffuse functions in the OBS used to solve the self-consistent field equations.
Jurgens-Lutovski and Alml\"of were the first to develop a correlated dual-basis method \cite{VRG:jurgens-lutovsky:1991:CPL} with the use of a smaller OBS and a larger VBS, reducing the cost of integral evaluation and transformation in the context of MP2 energy evaluation. Wolinski and Pulay later used the dual-basis approach to correct the basis set errors of {\em both} mean-field and correlation energies.\cite{VRG:wolinski:2003:JCP}
The dual-basis ideas have been revisited by many since then, both at the mean field \cite{VRG:lee:2000:CC,VRG:liang:2004:JPCA,VRG:nakajima:2006:JCP,VRG:tokura:2008:JCC,VRG:steele:2006:JPCA,VRG:deng:2009:JCP,VRG:maschio:2020:JCTC,VRG:maschio:2024:JPCA} and correlated (MP2\cite{VRG:steele:2006:JCP,VRG:distasio:2007:MP,VRG:steele:2009:JCTC,VRG:deng:2011:JCP} and MP2-F12\cite{VRG:valeev:2006:CPL}) methods. Modern explicitly-correlated F12 methods also use a dual-basis approach to correct for the basis set error of the reference mean-field energy, known as the CABS singles correction.\cite{VRG:adler:2007:JCP,VRG:knizia:2009:JCP} Although there are many nonessential differences in these approaches, one essential trait common to most dual basis approaches to date is that VBS spans OBS exactly (with some exceptions\cite{VRG:havriliak:1983:JACS, VRG:deng:2009:JCP,VRG:deng:2011:JCP,VRG:valeev:2006:CPL}). Secondly, the use of correlated dual-basis approaches has been limited to MP2, with only exceptions being the work of Zhang and Dolg\cite{VRG:zhang:2013:JCTC,VRG:zhang:2014:JCP,VRG:zhang:2015:JCTC} who leveraged dual-basis standard and explicitly-correlated CCSD(T) in the context of the incremental expansion, and Klopper and coworkers, who used a separate basis set for the computation of the perturbative triples [(T)] energy \cite{VRG:klopper:1997:TCA}. 

Our formulation of the dual-basis explicitly-correlated coupled-cluster method here largely builds on the original dual-basis MP2-F12 method developed in our group\cite{VRG:valeev:2006:CPL} and along the lines of the dual-basis CC methods of Zhang and Dolg,\cite{VRG:zhang:2013:JCTC,VRG:zhang:2014:JCP,VRG:zhang:2015:JCTC} except that (1) the dual-basis CC method is used to evaluate the correlation energy of the full system without domain decomposition, and (2) the perturbative CABS singles correction is reformulated to account for the significant magnitude of singles amplitudes induced by correlation and OBS$\to$VBS basis expansion.
The correlating (VBS) orbitals are constructed from the union of the OBS and VBS AO bases (VBS+ AO) as an orthogonal complement to the OBS orbitals; linear dependencies in the OBS, VBS, and VBS+ AO sets are eliminated by canonical orthogonalization. This process is equivalent to the CABS+ construction previously reported by our group. \cite{VRG:valeev:2004:CPL} Since we use Dunning's correlation consistent basis sets and their diffuse-augmented analogs (cc-pV$X$Z and aug-cc-pV$X$Z ($X$=D,T,Q,5)) \cite{VRG:dunning:1989:JCP, VRG:kendall:1992:JCP, VRG:woon:1993:JCP} for the OBS and VBS spaces respectively, the VBS+ and VBS AO bases are equivalent.

The use of different AO basis sets as support for reference and correlating orbitals has almost no effect on the CC-F12 formalism, provided that the Brillouin condition ($F^i_a=0$) is not assumed.
 In the context of dual-basis MP2 methods, 
violation of the Brillouin condition admits further relaxation of the reference orbitals which can be effected perturbatively via single excitations into VBS orbitals;\cite{ VRG:wolinski:2003:JCP, VRG:steele:2006:JCP, VRG:ksiazek:2008:MP} in the dual-basis CC framework these and correlation-induced orbital relaxation effects are already accounted for by the 1-body cluster operator:
\begin{align}
\label{eq:T1}
\hat{T}_1 \equiv t^i_a a^a_i.
\end{align}
The only modification to the CC-F12 formalism enters via the CABS singles correction which accounts for the basis set incompleteness of the reference by perturbative $i\to\alpha'$ singles replacements,\cite{VRG:adler:2007:JCP,VRG:knizia:2008:JCP} where $\alpha'$ represents an orbital in the CABS space but not in the OBS or VBS spaces. In the standard F12 methods this correction is obtained along the lines of the SCF energy correction in the dual-basis methods, which is appropriate for MP2-level treatments. In the dual-basis CC-F12 framework it is necessary to revise the definition of this correction to account for the presence of other 1-body cluster operators of significant magnitude.
To understand the requisite changes in the coupled-cluster F12 formalism, consider the CC Langrangian:
\begin{align}
\mathcal{L} = & \bra{0} (1+\hat{\Lambda}) \bar{H} \ket{0},
\end{align}
where $\bar{H} \equiv \exp(-\hat{T}) \hat{H} \exp(\hat{T}) = \hat{H} +[\hat{H}, \hat{T}] + \frac{1}{2!}[[\hat{H},\hat{T}],\hat{T}] \dots $.
For simplicity, let us limit ourselves to CCSD-F12, in which the cluster operator is defined as
\begin{align}
\label{eq:T}
\hat{T} \overset{\text{CCSD-F12}}{=} \hat{T}_1 + \hat{T}_2 + \hat{T}'_1 + \hat{T}'_2,
\end{align}
where $\hat{T}_1$ is the conventional 1-body cluster operator(\cref{eq:T1}), $\hat{T}_2$ is its 2-body counterpart,
\begin{align}
\label{eq:T2}
\hat{T}_2 \equiv \frac{1}{4} t^{ij}_{ab} a^{ab}_{ij},
\end{align}
$\hat{T}'_1$ accounts for orbital relaxation due to the mixing of OBS and CABS,
\begin{align}
\label{eq:t1prime}
\hat{T}'_1 \equiv t^m_{\alpha'} a^{\alpha'}_m,
\end{align}
(note the use of {\em all}, rather than just active, occupied states here)
and $\hat{T}'_2$ is defined as in \cref{eq:t2prime}.
The Langrange multiplier operator $\hat{\Lambda}$
is defined as the sum of Hermitian adjoints of \cref{eq:T1,eq:T2,eq:t1prime}, where the cluster amplitudes are replaced by the corresponding Lagrange multipliers $\lambda$. The CC amplitude equations define the stationarity conditions of the Lagrangian with respect to the multipliers:
\begin{align}
\label{eq:T1-eq}
\frac{\partial \mathcal{L}}{\partial \lambda^a_i} = & \bra{0} a^i_a \bar{H} \ket{0} = 0, \\
\label{eq:T2-eq}
\frac{\partial \mathcal{L}}{\partial \lambda^{ab}_{ij}} = & \bra{0} a^{ij}_{ab} \bar{H} \ket{0} = 0, \\
\label{eq:T1'-eq}
\frac{\partial \mathcal{L}}{\partial \lambda^{\alpha'}_m} = & \bra{0} a^m_{\alpha'} \bar{H} \ket{0} = 0.
\end{align}

Although the conventional singles and doubles equations (\cref{eq:T1-eq,eq:T2-eq}) include the contributions from the CABS singles amplitudes, when the exact\cite{VRG:shiozaki:2008:JCP,VRG:kohn:2008:JCP} and approximate\cite{VRG:fliegl:2005:JCP,VRG:adler:2007:JCP,VRG:valeev:2008:PCCPP,VRG:hattig:2010:JCP} CCSD-F12 methods are implemented in practice, the contributions of the CABS singles operator $\hat{T}'_1$ to the equations defining the conventional cluster operators have always been omitted.\footnote{\samepage Furthermore, when the F12 terms are incorporated perturbatively, as is done here via the \ccsdptf method,\cite{VRG:valeev:2008:PCCP, VRG:torheyden:2008:PCCP,VRG:valeev:2008:JCP,VRG:zhang:2012:JCTC} the geminal cluster operator $\hat{T}'_2$ is omitted from the conventional singles and doubles equations (\cref{eq:T1-eq,eq:T2-eq}).}
This can be rationalized by the fact that the basis set relaxation of the reference orbitals is a smaller effect than the many-body correlation effects within the fixed basis (hence the coupling of the corresponding amplitudes should be smaller than either).
Thus, traditional approaches that account for the effect of auxiliary (i.e. CABS) singles $\hat{T}_1'$ ignore its coupling to the conventional correlation cluster operators, that is, treat $\hat{T}_1'$ as purely correcting the basis set incompleteness of the Hartree-Fock reference.
For example, Wolinski and Pulay (WP) determined the auxiliary singles $t^m_{\alpha'}$ as a first-order perturbation in the context of the dual-basis MP2 approach\cite{VRG:wolinski:2003:JCP}: 
\begin{align}
    \frac{\partial \mathcal{L}}{\partial \lambda^{\alpha'}_m} \overset{\mathrm{WP}}{=} \bra{0} a^m_{\alpha'} \left(\hat{F} + [\hat{F}, \hat{T}'_1]\right) \ket{0} = F^m_{\alpha'} + F^{\beta'}_{\alpha'} t^m_{\beta'} - F^m_{n} t^n_{\alpha'} = 0,
\end{align}
and the second-order energy was obtained via $t^m_{\alpha'} F^{\alpha'}_m$.
In the context of the coupled-cluster F12 methods, Adler, Knizia and Werner (AKW)\cite{VRG:adler:2007:JCP} incorporated auxiliary singles using the first-order perturbative framework of WP, but with additional modifications. Namely, they recognized the importance of including the coupling between conventional and auxiliary singles, thus $\hat{T}_1'$ was extended to include excitations into both conventional and CABS unoccupied virtuals:
    \begin{align}
        \hat{T}_1' \overset{\mathrm{AKW}}{\equiv} & t^m_\alpha a^\alpha_m.
        \label{eq:t1prime-akw} 
    \end{align}
Of course, such a definition only makes sense in the a posteriori treatment of $\hat{T}_1'$, since including this form of \cref{eq:t1prime} alongside $\hat{T}_1$ is redundant.
The resulting AKW equations for first-order CABS singles,
\begin{align}
    \frac{\partial \mathcal{L}}{\partial \lambda^{\alpha}_m} \overset{\mathrm{AKW}}{=} \bra{0} a^m_{\alpha} \left(\hat{F} + [\hat{F}, \hat{T}'_1]\right) \ket{0} = F^m_{\alpha} + F^{\beta}_{\alpha} t^m_{\beta} - F^m_{n} t^n_{\alpha} = 0,
    \label{eq:apw}
\end{align}
and the second-order energy,
\begin{align}
    E^{(2)}_{\hat{T}_1'} \overset{\mathrm{AKW}}{=} t^m_{\alpha} F^{\alpha}_m,
    \label{eq:e2_apw}
\end{align}
are identical to the WP expressions, modulo substituting $\alpha' \to \alpha$. Note that for non-Brillouin references the AKW correction includes the conventional singles contribution to the MP2 energy \cite{VRG:knizia:2008:JCP}. Also note that in practical computations Werner et al. only use active occupied orbitals in the definition of \cref{eq:t1prime-akw}.

Considering the magnitude of the CC correlation and the loss of the Brillouin condition, the singles correction for the DB-CC-F12 formalism may need to be adjusted. Although the first-order perturbative treatment of the CABS singles is likely sufficient, the usual smallness of the conventional singles amplitudes cannot be assumed. Thus, instead of the Fock operators in \cref{eq:apw}, higher-order contributions to the CC $\bar{H}$ should be included. We have investigated two new variants of the CABS singles correction here, both of which include higher-order terms in $\bar{H}$. The first variant (PV0) defines $\hat{T}_1'$ as in \cref{eq:t1prime} whereas the other variant (PV1) defines $\hat{T}_1'$ as in the AKW approach (\cref{eq:t1prime-akw}).
To keep the cost of the CABS singles treatment negligible, we only include terms linear in $\hat{T}_1$ (since $\hat{T}_2$ is expected to be much smaller). The resulting amplitude equations are as follows:
\begin{subequations}
\begin{align}
\frac{\partial \mathcal{L}}{\partial \lambda^{\alpha'}_m} \overset{\mathrm{PV0}}{=} & \bra{0} a^m_{\alpha'} \left(\hat{F} + [\hat{H}, \hat{T}_1] + [\hat{F}, \hat{T}_1'] \right) \ket{0} \nonumber \\
= & F^m_{\alpha'} + \delta^m_k F^c_{\alpha'} t^k_c + \bar{g}^{mc}_{\alpha'k} t^k_c + F^{\beta'}_{\alpha'} t^m_{\beta'} - F^m_{n} t^n_{\alpha'}  ,
\label{eq:pv0}
\\
\frac{\partial \mathcal{L}}{\partial \lambda^{\alpha}_m} \overset{\mathrm{PV1}}{=} & \bra{0} a^m_{\alpha} \left(\hat{Q}'\left(\hat{F} + [\hat{H}, \hat{T}_1]\right) + [\hat{F}, \hat{T}_1'] \right) \ket{0} \nonumber \\
= & \delta^{\alpha'}_\alpha \left(F^m_{\alpha'} + \delta^m_k F^c_{\alpha'} t^k_c + \bar{g}^{mc}_{\alpha' k} t^k_c \right) + F^{\beta}_{\alpha} t^m_{\beta} - F^m_{n} t^n_{\alpha},
\label{eq:pv1}
\end{align}
\label{eq:pv}
\end{subequations}
with the $\hat{Q}'$ projector onto the CABS singles manifold:
\begin{align}
\hat{Q}' a^{\alpha'}_m \ket{0} \equiv & \, a^{\alpha'}_m \ket{0}, \\
\hat{Q}' a^a_m \ket{0} \equiv & \, 0.
\end{align}
The role of $\hat{Q}'$ in \cref{eq:pv1} is to avoid  double-counting of the conventional singles component of $\left(\hat{F} + [\hat{H},\hat{T}_1]\right)\ket{0}$, since they were already included in the evaluation of conventional 1-body amplitudes. Note that a similar double-counting of the conventional singles component of $\hat{F}\ket{0}$ in \cref{eq:apw} introduced the conventional singles MP2 energy contribution into the CABS singles correction in the AKW approach;
replacing $\hat{F}$ with $\hat{Q}'\hat{F}$ in  \cref{eq:apw} would eliminate it a priori, defining the AKW1 approach:
\begin{align}
    \frac{\partial \mathcal{L}}{\partial \lambda^{\alpha}_m} \overset{\mathrm{AKW1}}{=} \bra{0} a^m_{\alpha} \left( \hat{Q}' \hat{F} + [\hat{F}, \hat{T}'_1]\right) \ket{0} = \delta^{\alpha'}_\alpha F^m_{\alpha'} + F^{\beta}_{\alpha} t^m_{\beta} - F^m_{n} t^n_{\alpha} = 0,
    \label{eq:apw1}
\end{align}

The CABS singles energy expression in the dual-basis CCSD-F12 also differs from the AKW counterpart \cref{eq:e2_apw}. Since we included the effect of $\hat{T}_1$ on $\hat{T}_1'$ explicitly in \cref{eq:pv} it makes sense to account perturbatively for the effect of $\hat{T}_1'$ onto $\hat{T}_1$ by including the $\bra{0}\hat{\Lambda}_1 [\hat{H},\hat{T}_1']\ket{0}$ contribution to the Lagrangian; in the PV1 case this contribution must use $\hat{Q}' \hat{T}_1'$ instead of $\hat{T}_1'$ to avoid the double-counting.
Approximating $\hat{\Lambda}_1 \approx \hat{T}_1^\dagger$ results in the second-order CABS singles energy correction:

\begin{subequations}
\begin{align}
    E^{(2)}_{\hat{T}_1'} \overset{\mathrm{PV0}}{=} & \bra{0} \hat{T}_1^{\prime \dagger} \left(\hat{F} + [\hat{H}, \hat{T}_1] \right)  + \hat{T}_1^\dagger [\hat{H},\hat{T}_1'] \ket{0} \nonumber \\
    = & t_m^{\alpha'} F_{\alpha'}^m + 2 \Re t_m^{\alpha'} F^c_{\alpha'} t^k_c \delta^m_k + 2 \Re t_m^{\alpha'} \bar{g}^{mc}_{\alpha'k} t^k_c  \label{eq:e_pv0} \\
    E^{(2)}_{\hat{T}_1'} \overset{\mathrm{PV1}}{=} & \bra{0} \hat{T}_1^{\prime \dagger} \hat{Q}' \left(\hat{F} + [\hat{H}, \hat{T}_1] \right)  + \hat{T}_1^\dagger [\hat{H}, \hat{Q}' \hat{T}_1'] \ket{0} \nonumber \\
    = & t_m^{\alpha'} F_{\alpha'}^m + 2 \Re t_m^{\alpha'} F^c_{\alpha'} t^k_c \delta^m_k + 2 \Re t_m^{\alpha'} \bar{g}^{mc}_{\alpha' k} t^k_c \label{eq:e_pv1}
\end{align}
\label{eq:e_pv}
\end{subequations}
Note that the energy expressions here are identical; the difference lies in the determination of the CABS singles amplitudes ($t^{\alpha^\prime}_m$), which are computed differently for PV0 and PV1 (see \cref{eq:pv}).  The corresponding expression for the AKW1 approach is obtained analogously
\begin{align}
    E^{(2)}_{\hat{T}_1'} \overset{\mathrm{AKW1}}{=} 
    \bra{0} \hat{T}_1^{\prime \dagger} \hat{Q}' \hat{F}  \ket{0} = t^m_{\alpha'} F^{\alpha'}_m,
    \label{eq:e2_apw1}
\end{align}

The PV CABS formulations seem to improve the accuracy of chemical properties as we will discuss later, and are recommended for all CC-F12 calculations. The AKW approach is often sufficient for single-basis calculations, but the PV approaches are more rigorous and do not incur significantly higher computational cost. 

In any case, it is always recommended that the complete unoccupied space is used as the projection space rather than the CABS orbital space (i.e. PV1 over PV0 and AKW1 over AKW). The DB results are even more dependent on the more robust CABS results. 

\section{Technical Details}\label{sec:technical}

The dual-basis CC and CC-F12 approaches have been implemented in a developmental version of the Massively Parallel Quantum Chemistry (MPQC) package\cite{VRG:peng:2020:JCP}.
All F12 computations here used the \ccsdptf method\cite{VRG:valeev:2008:PCCP,VRG:valeev:2008:JCP} with perturbative inclusion of the geminal terms; for simplicity we will refer to them as CCSD-F12 (not to be confused with the complete CCSD-F12 method\cite{VRG:shiozaki:2008:PCCP,VRG:shiozaki:2008:JCP,VRG:kohn:2008:JCP}).

All calculations were performed with uncorrelated core electrons. For all dual-basis calculations presented here, the OBS was one of Dunning's correlation consistent cc-pV$X$Z ($X$=\{D,T,Q,5\}) basis sets (abbreviated $X$Z) and the corresponding VBS was the associated aug-cc-pV$X$Z basis (abbreviated a$X$Z). \cite{VRG:dunning:1989:JCP, VRG:kendall:1992:JCP,VRG:woon:1993:JCP,VRG:woon:1994:JCP} Single basis calculations were performed with both the $X$Z and a$X$Z basis sets. All 2-electron integrals were approximated using density fitting (DF) with the matching DF basis set: cc-pV$X$Z-RI for the single-basis $X$Z calculations, and aug-cc-pV$X$Z-RI for the dual-basis and single basis a$X$Z calculations. \cite{VRG:weigend:2002:JCP, VRG:hattig:2005:PCCP} 

The DB calculations must use the density fitting basis associated with the (larger) VBS to ensure that the errors due to DF are sufficiently small. This is the only difference between the HF portions of the $X$Z and DB $X$Z/a$X$Z calculations: the single-basis $X$Z calculations used $X$Z-RI for density fitting, while $X$Z/a$X$Z and a$X$Z calculations both used a$X$Z-RI. This difference causes minor disagreements on the order of DF errors (a few tens of microhartrees) between the HF energies of the two. For F12 calculations, the aug-cc-pV$X$Z/OptRI bases were used for the expansion of the geminal integrals.\cite{VRG:yousaf:2009:CPL} The standard Slater-type correlation factor,\cite{VRG:ten-no:2004:CPL} \cref{eq:f12}, was used. Most of the computations used the standard exponents recommended by Peterson et al.\cite{VRG:peterson:2008:JCP}: $\gamma=\{1.1,1.2,1.4,1.4\}$ for $X$=\{D,T,Q,5\}.
Some computations also used our revised values of $\gamma$ suited for CC-F12 calculations.\cite{VRG:powell:2025:JCTC}

Throughout this paper, we do not report individual component errors for the dual-basis calculations because the CCSD correlation energy, $E_\text{CCSD}^\text{corr}$ is not well-defined in that, if OBS $\neq$ VBS, $E_\text{CCSD}^\text{DB}$ contains conventional correlation energy, $E_\text{CCSD}^\text{corr}$, as well as a correction for the reference basis incompleteness relative to the VBS as described in \cref{sec:formalism}, and therefore it is not reasonable to compare component energies to the single-basis correlation energies. 
The obtained $X$Z/a$X$Z $E_\text{HF}$ are nearly identical to the single-basis $X$Z values, as described above. The DB approach functions by way of an error cancellation  effect in which the correlation energy of the dual-basis calculation is larger in magnitude even than the a$X$Z correlation energy because it includes the additional energetic effects of relaxing the SCF orbitals into the larger VBS basis. The dual-basis CC energies, therefore, cannot be meaningfully partitioned into reference and correlation components. 

A set of \{ \ce{CH4}, \ce{NH3}, \ce{H2O}, \ce{HF}, \ce{Ne} \} was used to study absolute electronic energies, and the HJO12 set of 12 isogyric small-molecule reactions \cite{VRG:helgaker:2000:} was used to examine performance for chemical properties. The CBS HF and valence CCSD correlation energies for these molecules were estimated using the cc-pCV6Z HF energy and the aug-cc-pV\{Q,5\}Z CCSD energies extrapolated via the $X^{-3}$ scheme,\cite{VRG:helgaker:1997:JCP} respectively (as listed in \citenum{VRG:helgaker:2000:}). The S66 benchmark set\cite{VRG:rezac:2011:JCTC,VRG:brauer:2016:PCCP} was used to assess the applicability of the dual-basis approach to noncovalent interactions. 
The HF and CCSD CBS limits for the S66 set were taken from the ``silver'' benchmark of Brauer et al.\cite{VRG:brauer:2016:PCCP}

\section{Results}\label{sec:results}

\subsection{Absolute electronic energies}\label{sec:results:abs-elec}

\begin{figure}[htbp!]
    \centering
    \begin{subfigure}{0.35\textwidth}
        \centering
            \includegraphics[width=\textwidth]{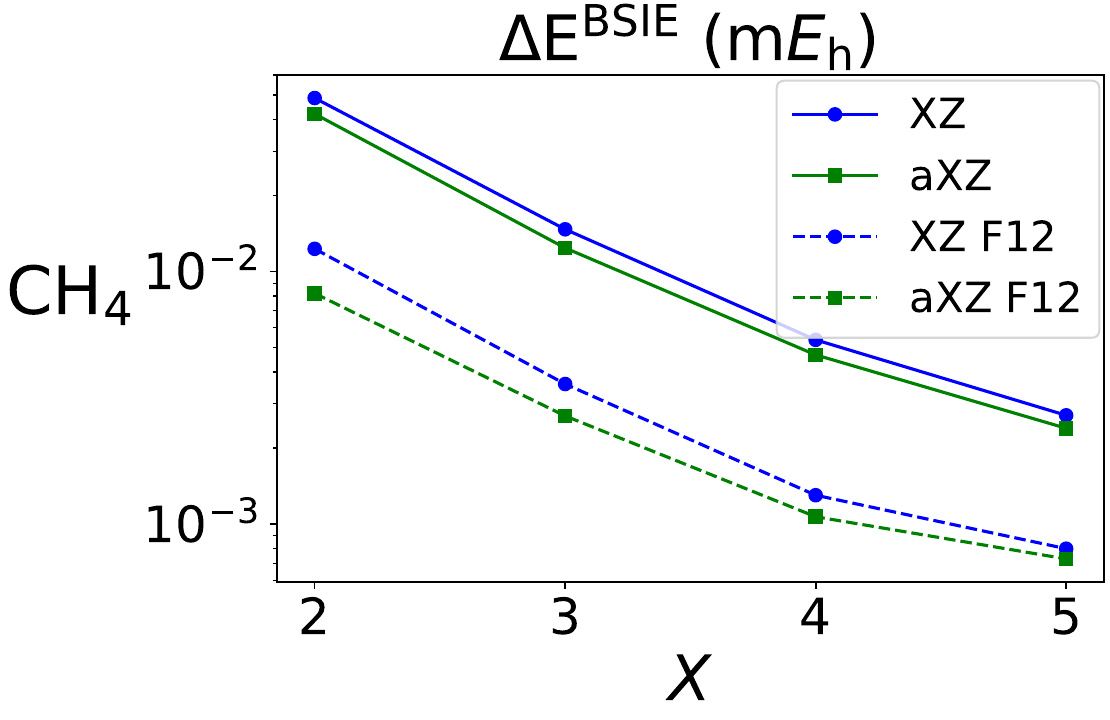}
        \label{fig:abs_e:ch4-bsie}
    \end{subfigure}
    \begin{subfigure}{0.3\textwidth}
        \centering
        \includegraphics[width=\textwidth]{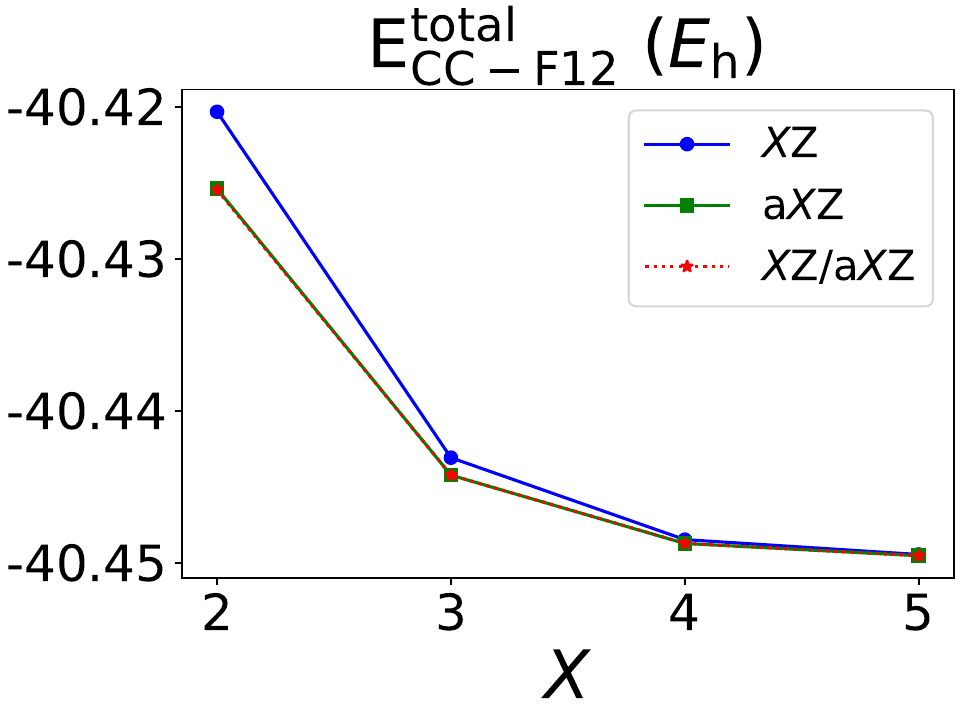}
        \label{fig:abs_e:ch4-ccf12}
    \end{subfigure}
    \\
    \begin{subfigure}{0.35\textwidth}
        \centering
        \includegraphics[width=\textwidth]{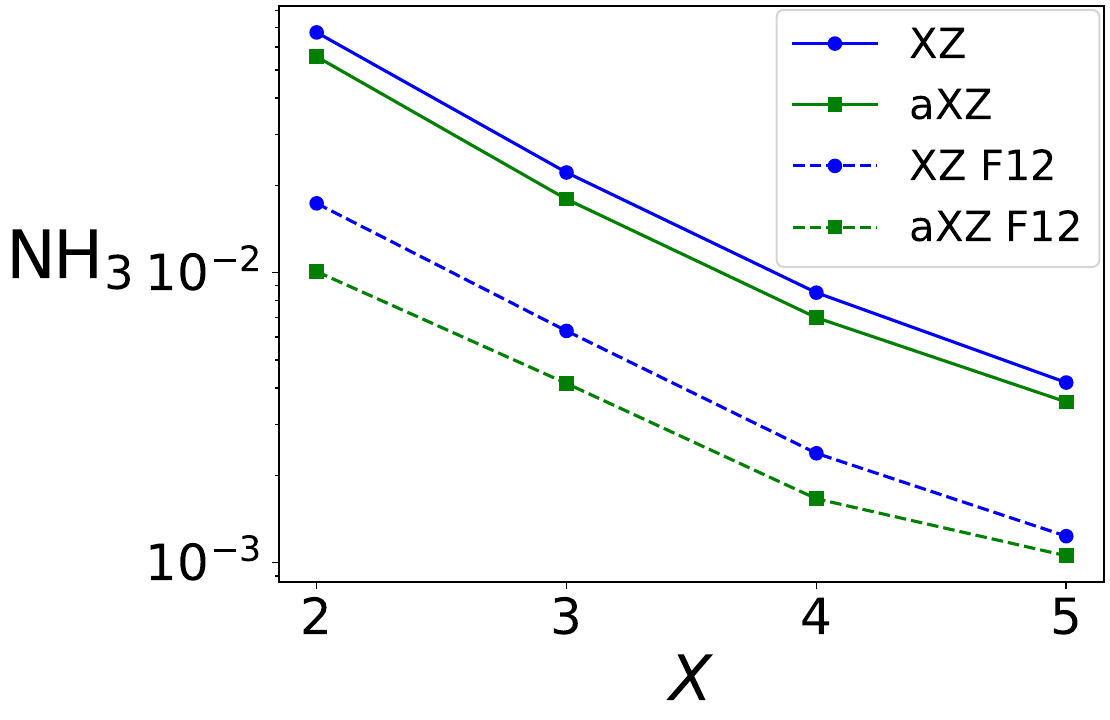}
        \label{fig:abs_e:nh3-bsie}
    \end{subfigure}
    \begin{subfigure}{0.3\textwidth}
        \centering
        \includegraphics[width=\textwidth]{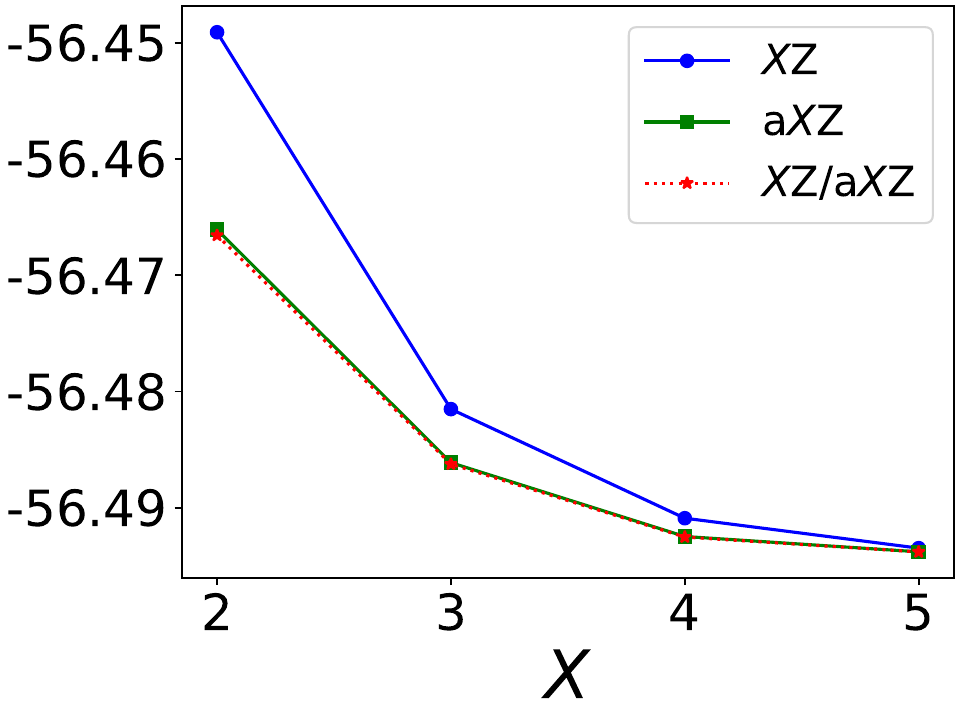}
        \label{fig:abs_e:nh3-ccf12}
    \end{subfigure}
    \\
    \begin{subfigure}{0.35\textwidth}
        \centering
        \includegraphics[width=\textwidth]{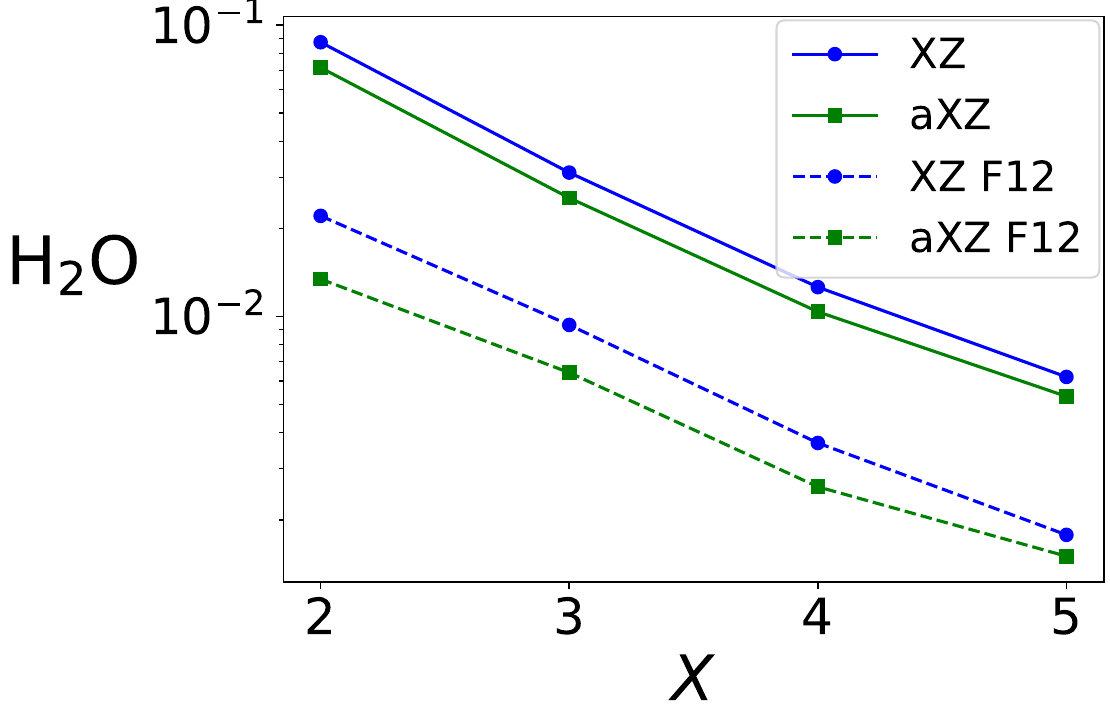}
        \label{fig:abs_e:h2o-bsie}
    \end{subfigure}
    \begin{subfigure}{0.3\textwidth}
        \centering
        \includegraphics[width=\textwidth]{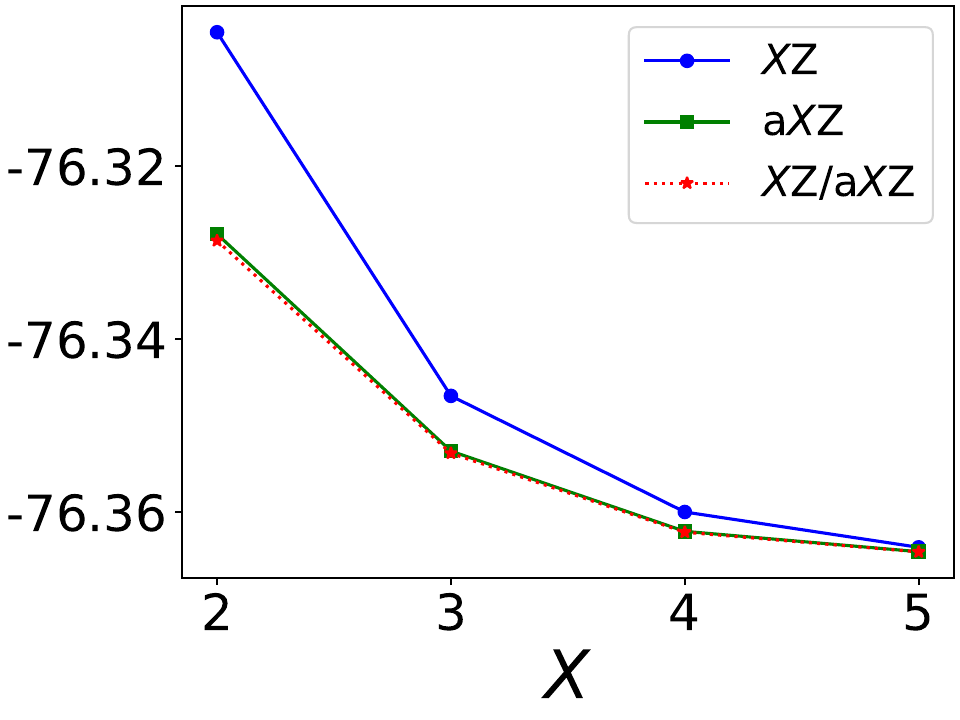}
        \label{fig:abs_e:h2o-ccf12}
    \end{subfigure}
    \\
    \begin{subfigure}{0.35\textwidth}
        \centering
        \includegraphics[width=\textwidth]{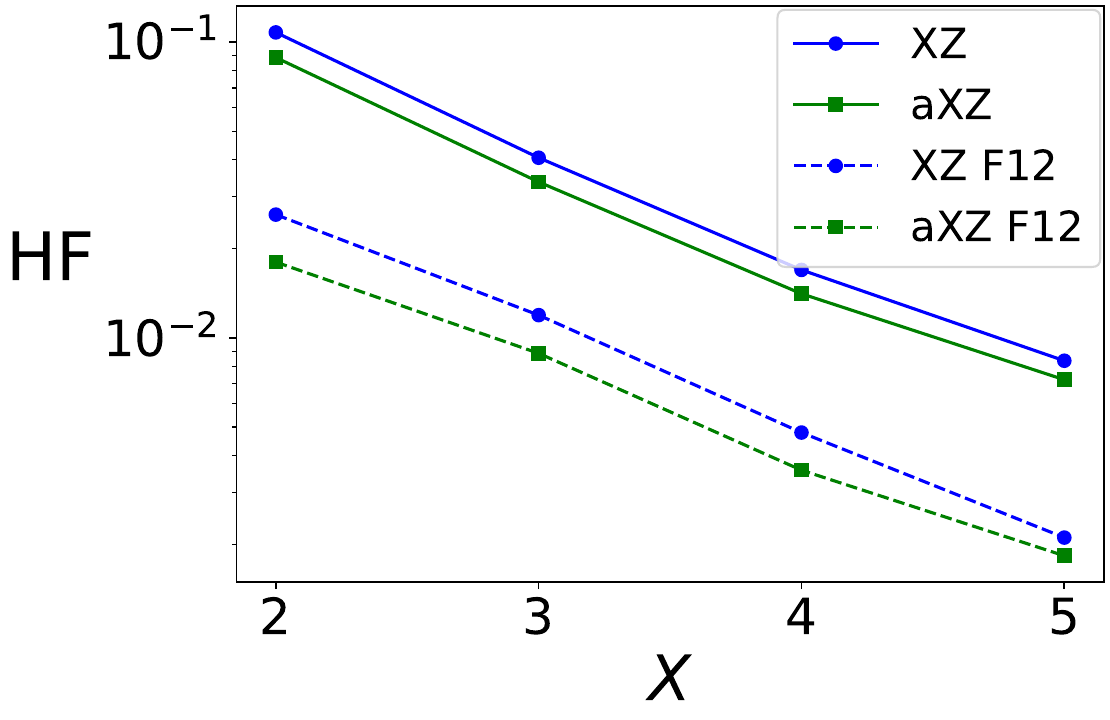}
        \label{fig:abs_e:hf-bsie}
    \end{subfigure}
    \begin{subfigure}{0.3\textwidth}
        \centering
        \includegraphics[width=\textwidth]{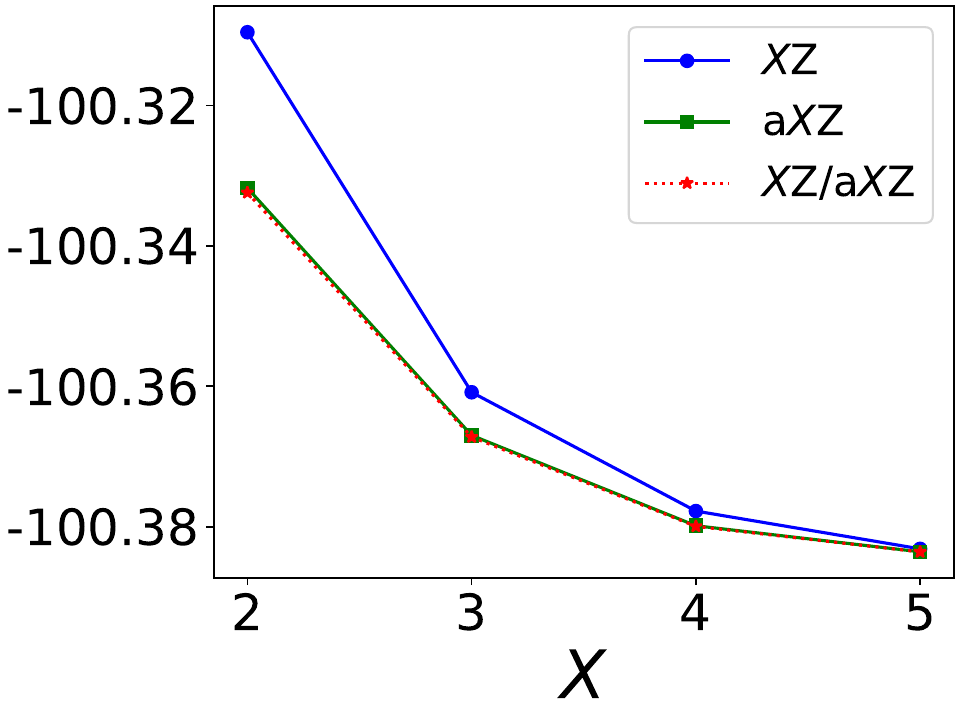}
        \label{fig:abs_e:hf-ccf12}
    \end{subfigure}
    \\
    \begin{subfigure}{0.35\textwidth}
        \centering
        \includegraphics[width=\textwidth]{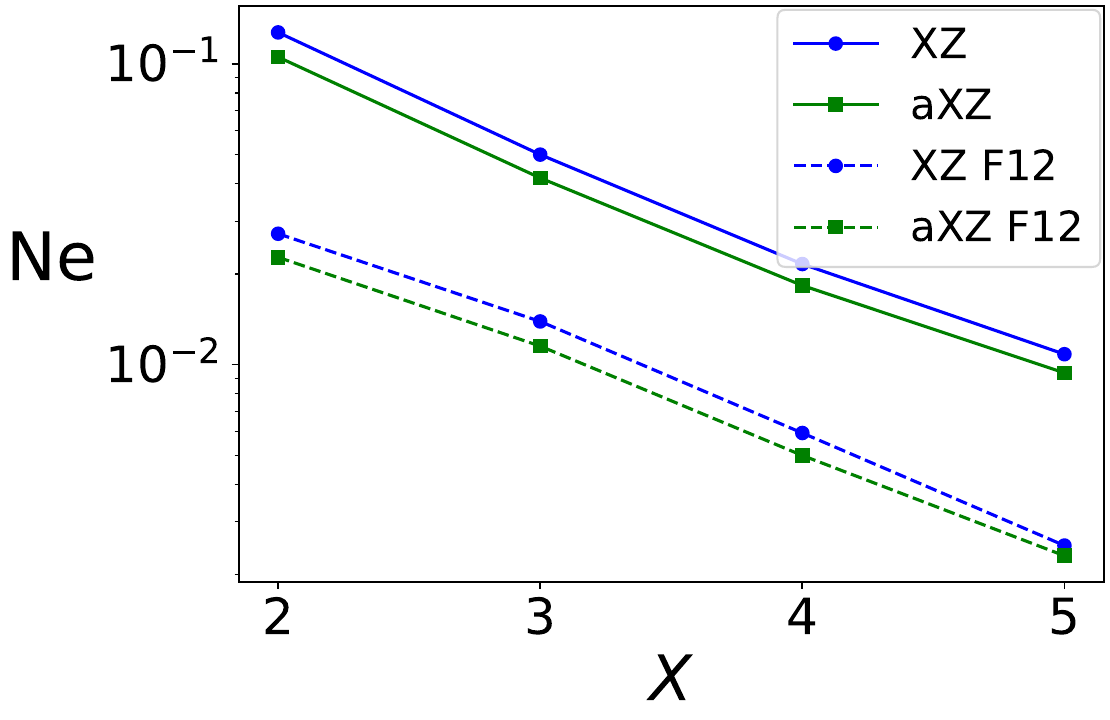}
        \label{fig:abs_e:ne-bsie}
    \end{subfigure}
    \begin{subfigure}{0.3\textwidth}
        \centering
        \includegraphics[width=\textwidth]{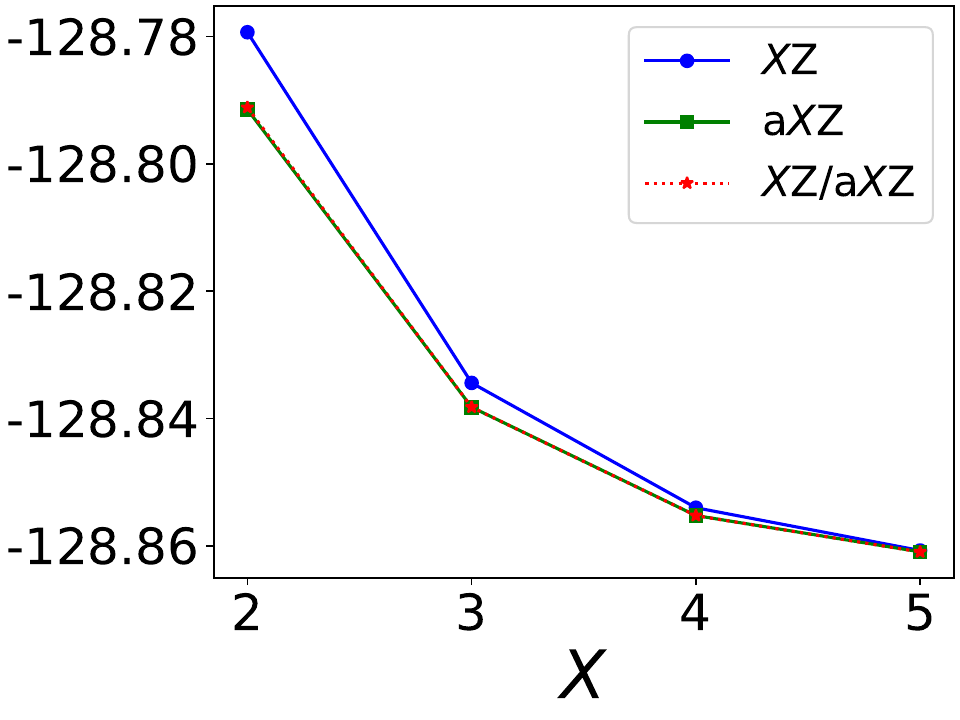}
        \label{fig:abs_e:ne-ccf12}
    \end{subfigure}
                \\
    \caption{The basis set convergence of CCSD and CCSD-F12 energies of the 10-electron benchmark set with and without diffuse AOs.} 
    \label{fig:abs_e}
\end{figure}

First we consider the impact of diffuse AOs on absolute electronic energies, using the series \{CH$_4$, NH$_3$, H$_2$O, HF, Ne\} of paradigmatic 10-electron neutral species as an example.
Although the absolute effect of adding diffuse functions decreases with the cardinal number $X$, in the relative sense, their effect varies with $X$ only weakly.
This is easily seen in the left panel of \cref{fig:abs_e}: the addition of diffuse functions translates into a proportional increase in the CCSD correlation energy roughly independent of $X$, so the $X$Z and a$X$Z CCSD lines are roughly parallel to each other.

We quantified the absolute and relative energy lowering due to the inclusion of diffuse AOs by
\begin{align}
    \Delta E^\text{aug}_\text{XZ} \equiv E_\text{a$X$Z} - E_\text{$X$Z}
\end{align}
and 
\begin{align}
    R^\text{aug}_\text{XZ} \equiv \frac{\Delta E^\text{aug}_\text{XZ}}{\Delta E_\text{$X$Z}^\text{BSIE}} ,
    \label{eq:Raug}
\end{align}
respectively,
where
\begin{align}
    \Delta E^\text{BSIE}_\text{OBS} \equiv E_\text{CBS} - E_\text{OBS}
\end{align}
is the BSIE for the given OBS.
These quantities were gauged against the absolute and relative lowerings due to increasing the cardinal number by 1:
\begin{align}
    \Delta E^{X+1}_\text{$X$Z} \equiv & E_\text{$(X+1)$Z} - E_\text{$X$Z} \\
    R^{X+1}_\text{$X$Z} \equiv & \frac{\Delta E^{X+1}_\text{$X$Z}}{\Delta E_\text{$X$Z}^\text{BSIE}}
    \label{eq:RXp1}
\end{align}
The comparison revealed the following trends.

\subsubsection{$R^\mathrm{aug}$ vs $R^{X+1}$}
\label{sec:results:abs-elec:Raug-vs-RXp1}

\begin{figure}[ht!]
    \centering
    \begin{subfigure}{0.32\textwidth}
        \includegraphics[width=\textwidth]{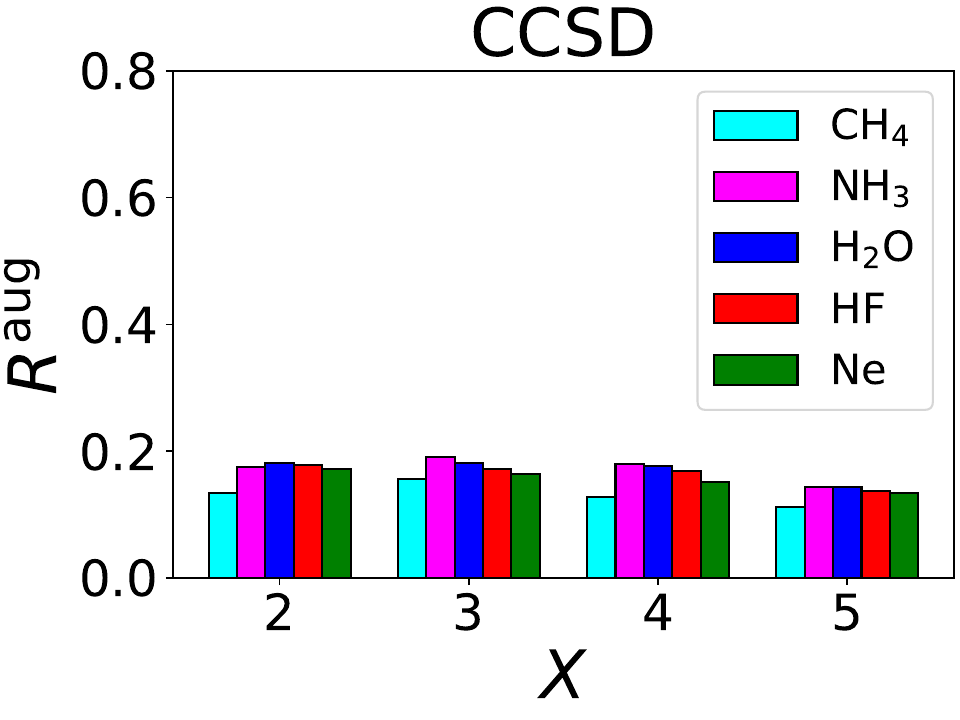}
        \caption{}
        \label{fig:Raug_ccsd}
    \end{subfigure}
    \begin{subfigure}{0.32\textwidth}
        \includegraphics[width=\textwidth]{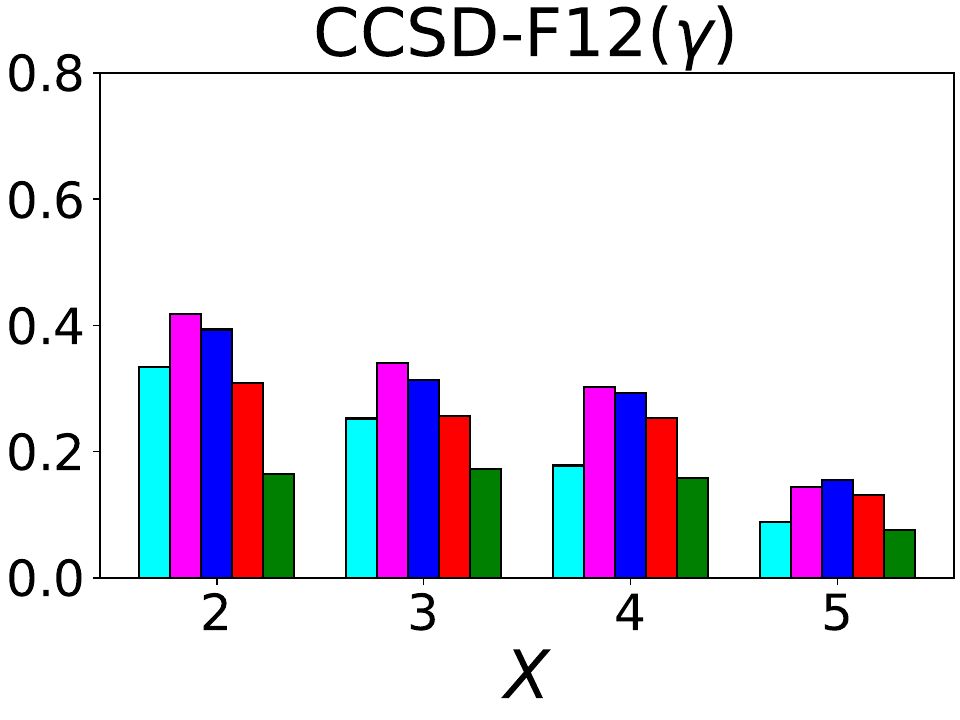}
        \caption{}
        \label{fig:Raug_ccsdf12}
    \end{subfigure} 
    \begin{subfigure}{0.32\textwidth}
        \includegraphics[width=\textwidth]{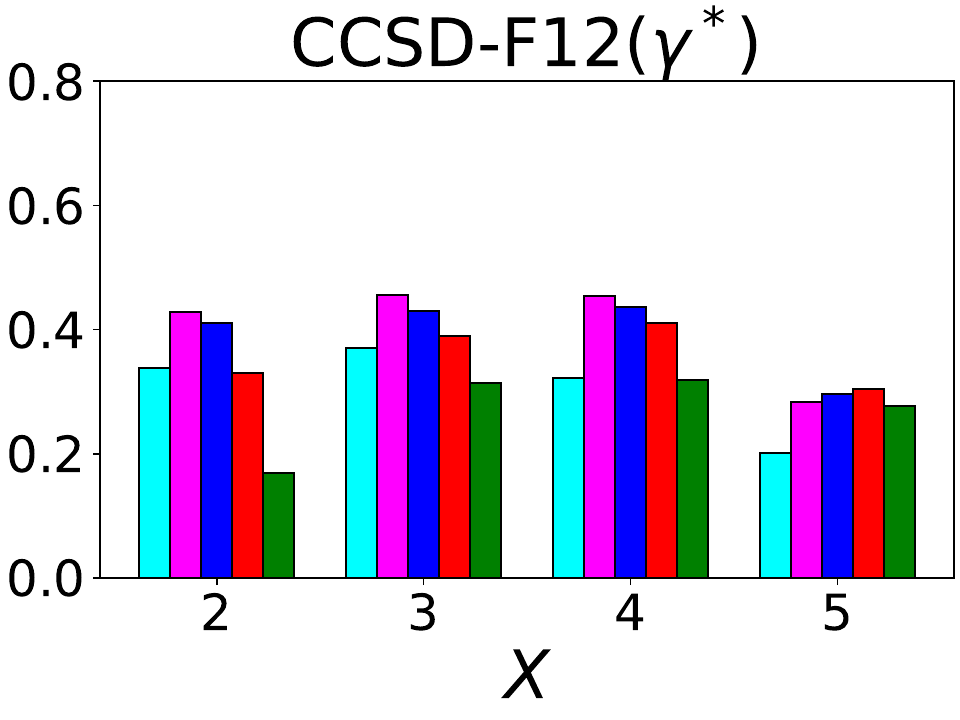}
        \caption{}
        \label{fig:Raug_ccsdf12_bo}
    \end{subfigure}\\
    \centering
    \begin{subfigure}{0.32\textwidth}
        \includegraphics[width=\textwidth]{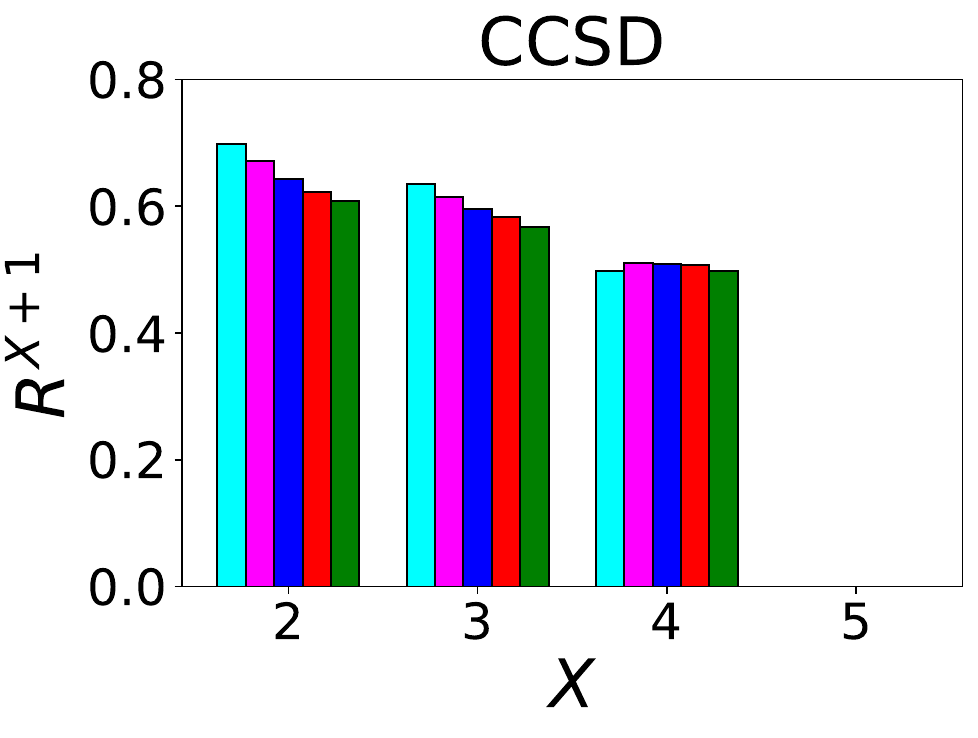}
        \caption{} 
        \label{fig:Rx_ccsd}
    \end{subfigure}
    \begin{subfigure}{0.32\textwidth}
        \includegraphics[width=\textwidth]{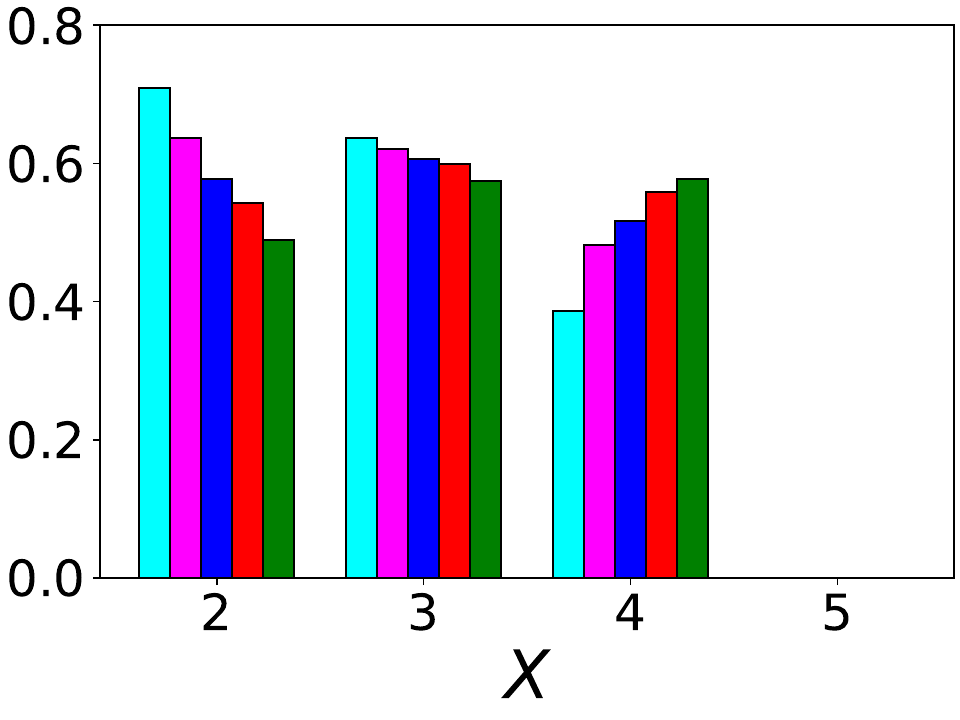}
        \caption{}
        \label{fig:Rx_ccsdf12}
    \end{subfigure}
    \begin{subfigure}{0.32\textwidth}
        \includegraphics[width=\textwidth]{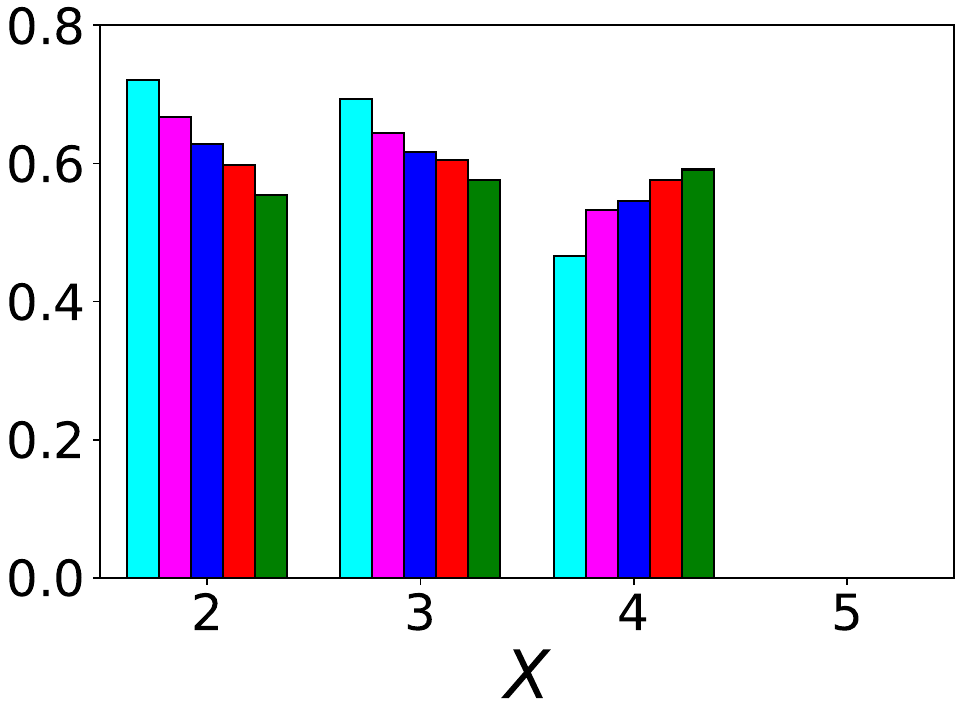}
        \caption{}
        \label{fig:Rx_ccsd_bo}
    \end{subfigure} \\
    \begin{subfigure}{0.32\textwidth}
        \includegraphics[width=\textwidth]{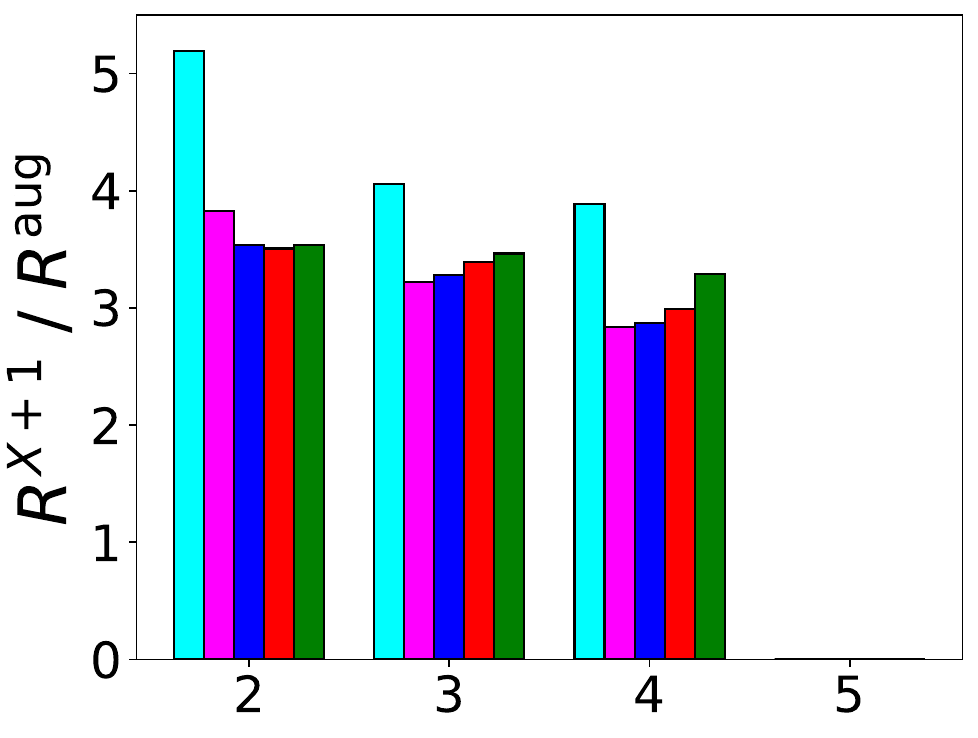}
        \caption{}
        \label{fig:RxRa_ccsd}
    \end{subfigure}
    \begin{subfigure}{0.32\textwidth}
        \includegraphics[width=\textwidth]{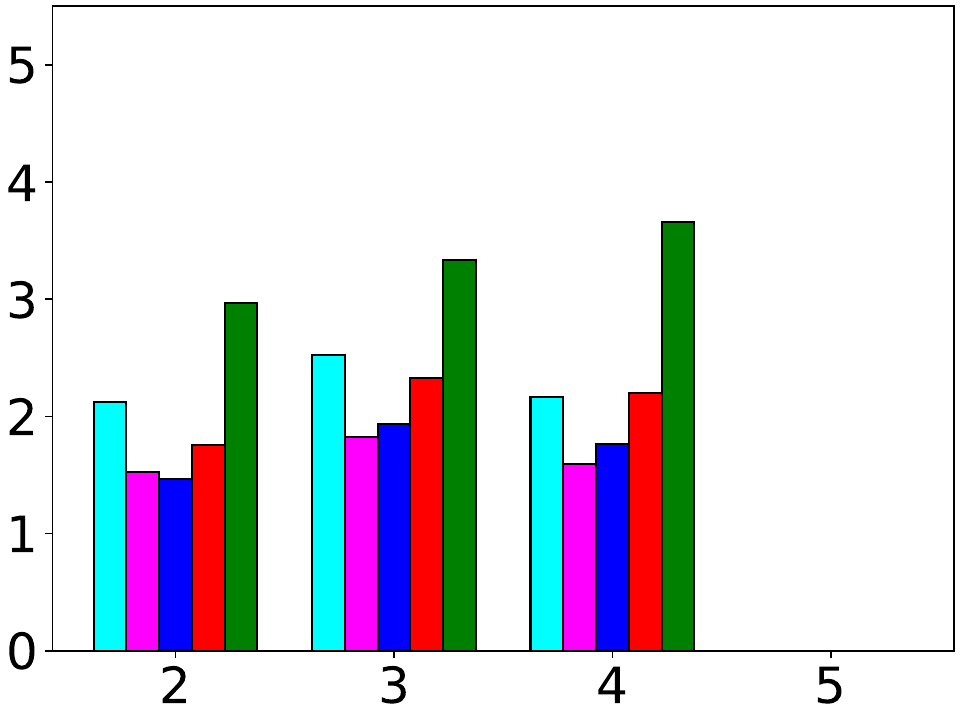}
        \caption{}
        \label{fig:RxRa_ccsdf12}
    \end{subfigure}
    \begin{subfigure}{0.32\textwidth}
        \includegraphics[width=\textwidth]{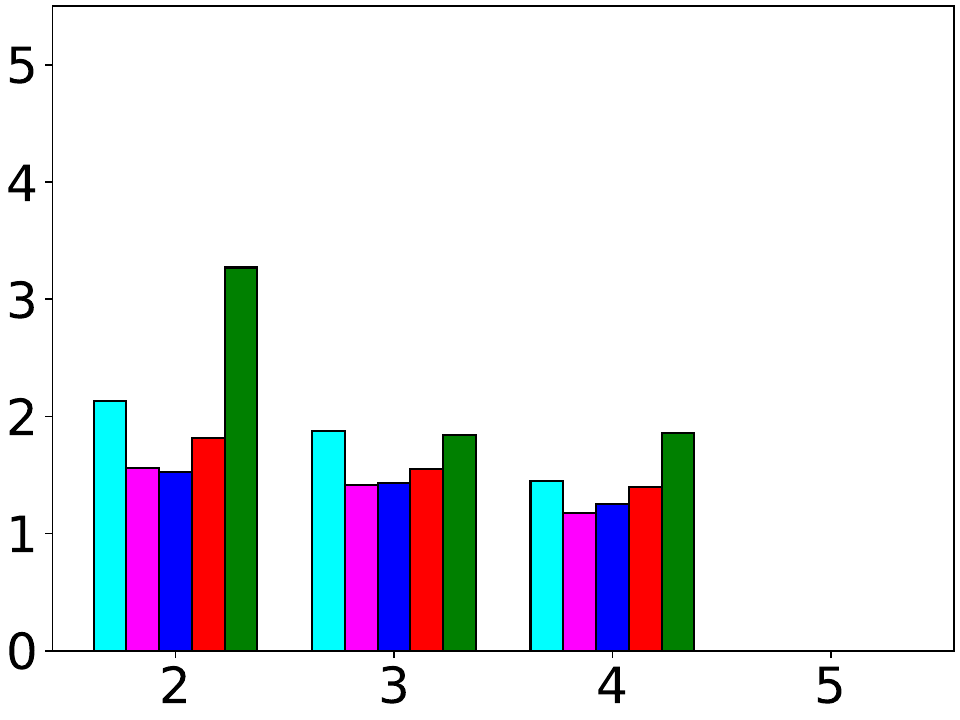}
        \caption{}
        \label{fig:RxRa_ccsdf12_bo}
    \end{subfigure}
    \caption{Relative impact of adding diffuse AOs [$R^\text{aug}$ (\cref{eq:Raug}), row 1] vs incrementing the OBS cardinal number [$R^{X+1}$ (\cref{eq:RXp1}), row 2] on CCSD correlation energy BSIE, without (column 1) and with F12 terms, using standard\cite{VRG:peterson:2008:JCP} (column 2) and revised\cite{VRG:powell:2025:JCTC} (column 3) geminal exponents.}
    \label{fig:bsie-barchart}
\end{figure}

{\bf CCSD}. The narrow range of values of $R^\text{aug}$ for {\em standard} (non-F12) CCSD,  $11\%$\textendash$19\%$, indicates that the addition of diffuse AOs changes the correlation energy by a proportion roughly independent of $X$ and the system type (\cref{fig:Raug_ccsd}). The energy lowering due to the addition of diffuse AOs is considerably smaller than that achieved by increasing the cardinal number of the basis: $R^{X+1}$ is $43\%$\textendash$70\%$ (\cref{fig:Rx_ccsd}).
In other words, incrementing the cardinal number is 3-4 times more important than adding diffuse functions for these neutral systems, as illustrated clearly by the $R^{X+1}/R^\text{aug}$ ratio (\cref{fig:RxRa_ccsd}).
This is not surprising, as the addition of diffuse functions to OBSs designed for correlated computations was motivated by the need to model anions, rather than neutral species\cite{VRG:kendall:1992:JCP}.

{\bf CCSD-F12}. $R^\text{aug}$ for the \emph{explicitly-correlated} CCSD calculations is far less systematic, both with respect to the system and $X$ (see \cref{fig:Raug_ccsdf12}). 
For the Ne atom, the values of CCSD-F12 $R^\text{aug}$ are similar to the CCSD counterparts for all $X$.
 However, for molecules, $R^\text{aug}$ for CCSD-F12 significantly exceeds that of the standard method, except for $X=5$. This is in sharp contrast to $R^{X+1}$, which does not differ significantly between CCSD-F12 (\cref{fig:Rx_ccsdf12}) and CCSD (\cref{fig:Rx_ccsd}).
As a result, the ratios $R^{X+1}/R^\text{aug}$ for CCSD-F12 (\cref{fig:RxRa_ccsdf12}) are significantly smaller than those of CCSD (\cref{fig:RxRa_ccsd}), namely $1.5$\textendash$2$ rather than $3$\textendash$4$.

{\bf F12 geminal exponent: standard vs revised}. During the course of this work we realized\cite{VRG:powell:2025:JCTC} that the optimal geminal exponents for the CC-F12 methods for a$X$Z OBS ($X\geq 3$) were substantially larger than the recommended values\cite{VRG:peterson:2008:JCP} obtained with the MP2-F12 method. Using the revised exponents (the right column of \cref{fig:bsie-barchart}) increases $R^\text{aug}$ substantially without much change in $R^{X+1}$. As a result, the ratio $R^{X+1}/R^\text{aug}$ decreases even further for $X \geq 3$, with most values substantially below 1.5 (\cref{fig:RxRa_ccsdf12_bo}) compared to $\sim 3$ for CCSD (\cref{fig:RxRa_ccsd}).

\subsubsection{$\Delta E^\text{aug}$: CC vs CC-F12}\label{sec:results:abs-elec:DEaug}

\begin{figure}[ht!]
    \centering
    \begin{subfigure}{0.32\textwidth}
        \includegraphics[width=\textwidth]{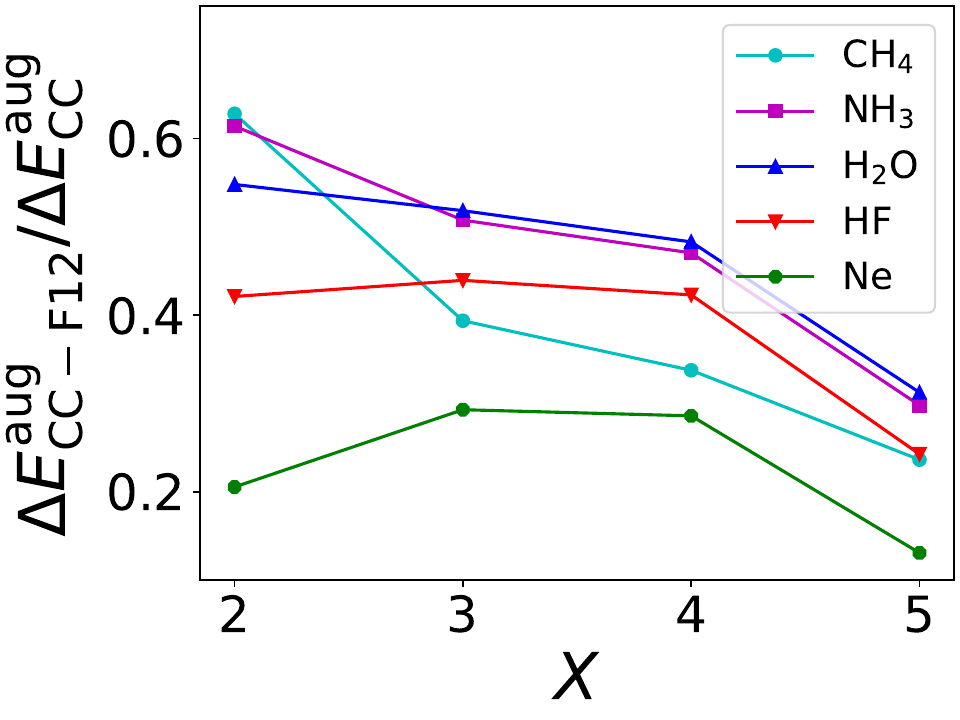}
        \caption{Standard (Ref. \citenum{VRG:peterson:2008:JCP}).}
        \label{fig:abs_e:ratio-diffuse-stdbeta}
    \end{subfigure}
    \hfill
    \begin{subfigure}{0.32\textwidth}
        \centering
        \includegraphics[width=\textwidth]{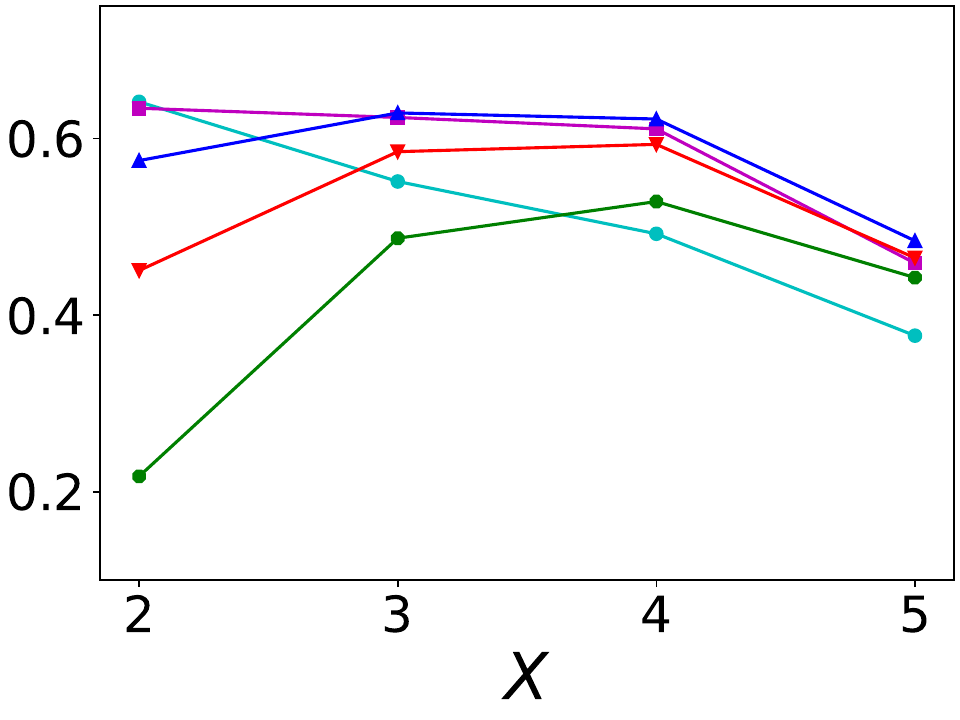}
        \caption{Revised (Ref. \citenum{VRG:powell:2025:JCTC}).}
        \label{fig:abs_e:ratio-diffuse-optbeta}
    \end{subfigure}
    \hfill
    \begin{subfigure}{0.32\textwidth}
        \centering
        \includegraphics[width=\textwidth]{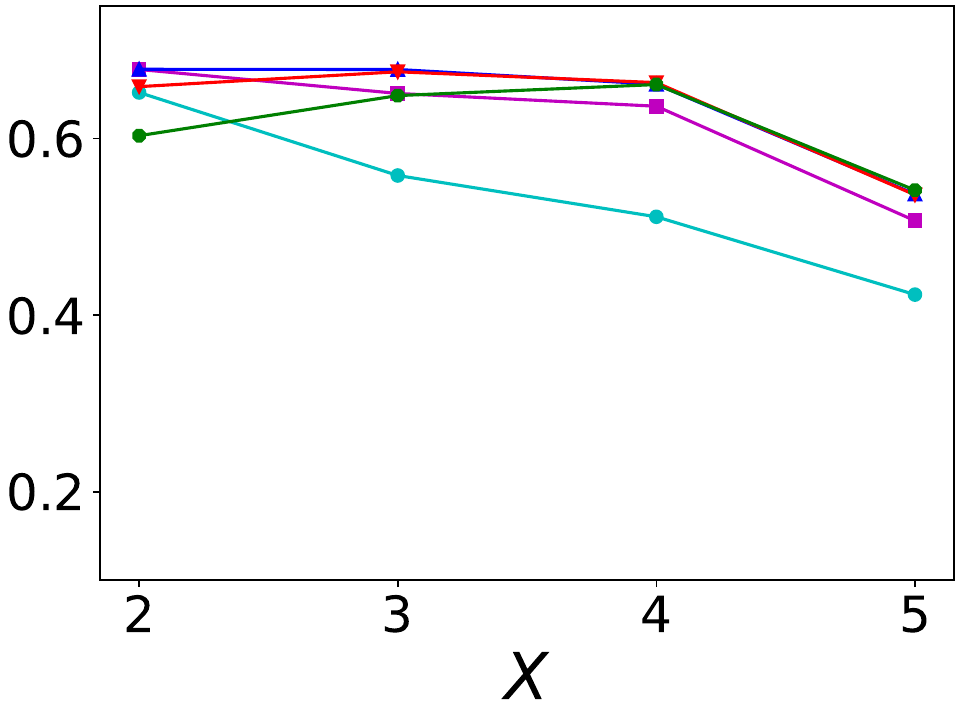}
        \caption{System-optimized (Ref. \citenum{VRG:powell:2025:JCTC}).}
        \label{fig:abs_e:ratio-diffuse-moloptbeta}
    \end{subfigure}
    \caption{Impact of the addition of diffuse AOs on CCSD-F12 correlation energy with three choices of geminal lengthscales (see text) relative to that of the CCSD counterpart.}
    \label{fig:abs_e:ratio-diffuse}
\end{figure}

The increased \emph{relative} importance of diffuse AOs for explicitly-correlated CC correlation energies, manifested as an increase of $R^\text{aug}$ and $R^{X+1}/R^\text{aug}$, does not imply that their \emph{absolute} impact on the CC-F12 correlation energy is greater than on the CCSD counterpart. As the anatomical analysis of the F12 ``amplitudes'' at the beginning of \cref{sec:formalism} revealed, geminals can partially account for the diffuse AO contributions to the virtual orbitals. To quantify this we examined the ratio of CCSD-F12 $\Delta E^\text{aug}$ to its non-F12 counterpart (\cref{fig:abs_e:ratio-diffuse}). This ratio becomes zero when the inclusion of the F12 terms perfectly compensates for the lack of diffuse functions in OBS. The higher the value of this ratio, the less capable the F12 terms are of such compensation. That the ratios are smaller than 1 indicates that the absolute impact of diffuse AOs on the CCSD-F12 correlation energy is reduced relative to that of the standard CCSD, but nevertheless, the F12 terms can describe the long-range correlations of electrons ordinarily described by the addition of diffuse AOs. The ratios are spread widely (0.1\textendash0.7) if standard geminal exponents are used (\cref{fig:abs_e:ratio-diffuse-stdbeta}). Using revised exponents (\cref{fig:abs_e:ratio-diffuse-optbeta}) increases the ratios and reduces their spread significantly. To push the envelope further, we optimized the geminal parameters for each system  individually using the optimization framework of Ref. \citenum{VRG:powell:2025:JCTC}  by minimizing the basis set incompleteness of the correlation energy. Since the Slater-type correlation factor has a nearly optimal shape\cite{VRG:johnson:2017:CPL} this can be viewed as the best-case scenario for the F12 methods without resorting to multiple correlation factors\cite{VRG:valeev:2006:JCP}. With these system-specific optimal geminal parameters the ratios increase even further (\cref{fig:abs_e:ratio-diffuse-moloptbeta}) and the resulting spread of the ratios is very narrow, $0.5$\textendash$0.6$, and varies weakly with $X$ for $2\leq X \leq 4$.  What is most remarkable is that optimizing the geminals by minimizing the BSIE reduces the compensation between the correlation energy contributions of F12 and diffuse AOs. This strongly suggests that the physics of correlations described by the geminals and the diffuse AOs are indeed distinct from each other.

Overall, with the use of the revised geminal exponents tuned for CC-F12 (while ignoring the Ne atom as an outlier), the effect of diffuse AOs on the CCSD-F12 correlation energy is roughly half of that of CCSD. However, the reduction in the correlation energy impact still translates into an \emph{increase} by a factor of $\sim$2  of the relative importance of diffuse AOs for CCSD-F12 energy, due to a significantly smaller BSIE of CCSD-F12 compared to that of CCSD. The increased importance of diffuse AOs thus seems to be due to the dramatic reduction of BSIE by the F12 terms, and thus the increased relative importance of the residual long-range correlations of electrons that are best modeled by diffuse AOs.

\subsubsection{Dual-Basis CC-F12}

As we discussed in the beginning of \cref{sec:formalism}, the diffuse AOs also affect the explicitly-correlated terms by changing the occupied orbitals. Although this effect should not be significant, we used the dual-basis CCSD-F12 method to verify this hypothesis.
Note that the DB-CCSD-F12 energies reported in this section used the AKW CABS singles correction for simplicity, since the more elaborate PV CABS approaches are mostly relevant for relative energies as we will examine in \cref{sec:results:hjo12,sec:results:s66}).

\begin{figure}
    \centering
    \begin{subfigure}{0.32\textwidth}
        \includegraphics[width=\linewidth]{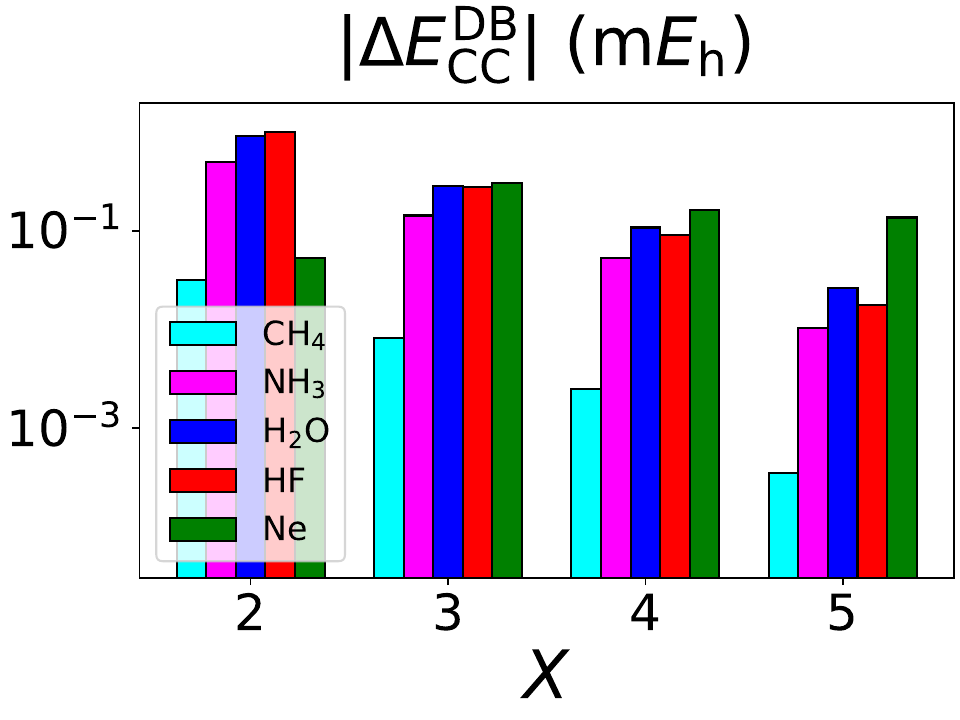}
        \caption{}
    \end{subfigure}
    \begin{subfigure}{0.32\textwidth}
        \includegraphics[width=\linewidth]{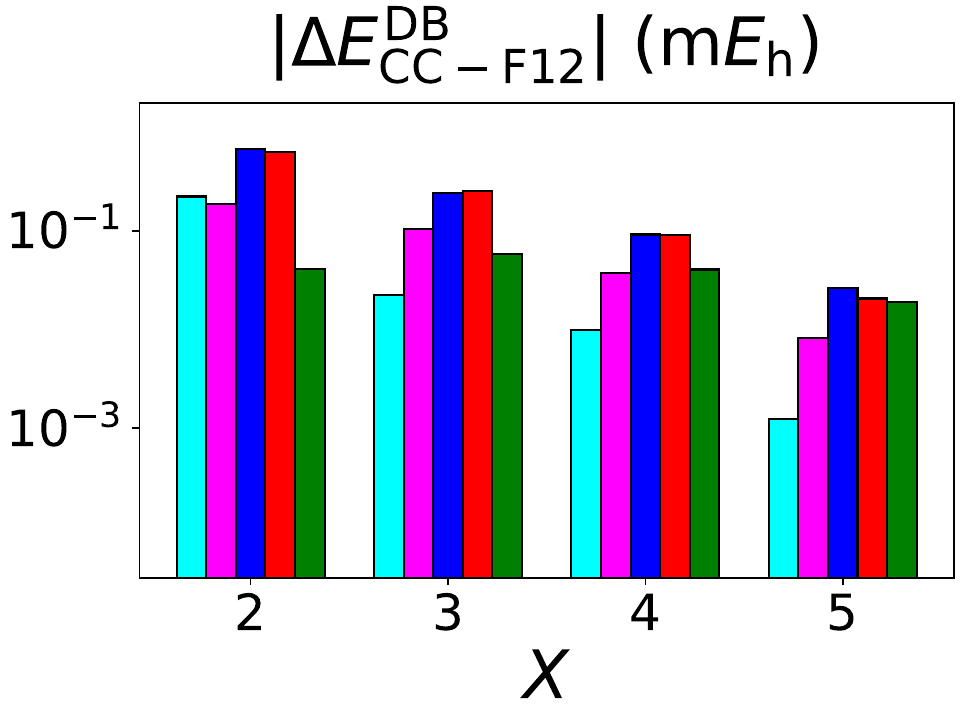}
        \caption{}
    \end{subfigure}
    \begin{subfigure}{0.32\textwidth}
        \includegraphics[width=\linewidth]{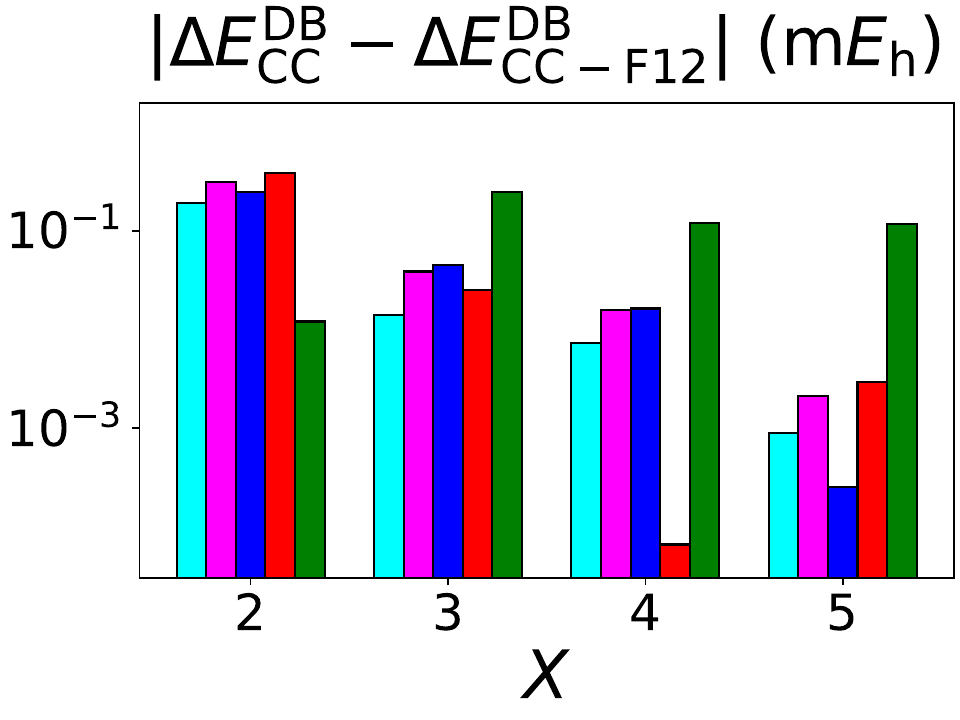}
        \caption{}
    \end{subfigure}
    \caption{Magnitudes of $\Delta E^\text{DB}$ for CCSD and CCSD-F12 total energies, and the respective differences ($mE_\text{h}$).}
    \label{fig:delta-db}
\end{figure}

We measured the difference between the total energy obtained with the standard a$X$Z OBS and its dual-basis $X$Z/a$X$Z counterpart:
\begin{equation}
    \Delta E^\text{DB} \equiv E^\text{a$X$Z} - E^\text{$X$Z/a$X$Z}
    \label{eq:DeltaEDB}
\end{equation}
As the plots of the total energies in the right column of \cref{fig:abs_e} illustrate, the CCSD-F12 $\Delta E^\text{DB}$ is negligible relative to $\Delta E^\text{aug}$. As illustrated in more detail in \cref{fig:delta-db}, $\Delta E^\text{DB}$ for both CCSD and CCSD-F12 does not exceed 1 $mE_\text{h}$ even for a double-zeta basis and decays rapidly with $X$. More importantly, the difference between CCSD and CCSD-F12 $\Delta E^\text{DB}$ is still smaller, on the order of a few tenths of a millihartree. Although it is not possible to decompose this difference into the contribution from the CABS singles and F12 terms cleanly, its magnitude can be viewed as an estimate of the change in the CABS singles and F12 energy corrections induced by the inclusion of diffuse AOs in the occupied orbitals. Clearly, this effect is much smaller than the corresponding effect of diffuse AOs on the total CCSD-F12 energy (\cref{fig:abs_e}). This is a strong indication that the bulk of the latter is due to the effect of diffuse AOs on virtual (correlating) orbitals, both directly via conventional correlation amplitudes and indirectly via the modification of the F12 amplitudes.

\subsection{Chemical Reaction Energies}\label{sec:results:hjo12}

\begin{table}[]
\centering
\begin{tabular}{c|cc|cc|ccc}
\hline \hline
 & \multicolumn{2}{c|}{HF$^a$} & \multicolumn{2}{c|}{correlation} & \multicolumn{3}{c}{total} \\
$X$ & $X$Z & a$X$Z & $X$Z & a$X$Z & $X$Z & a$X$Z & $X$Z/a$X$Z \\ \hline
 & \multicolumn{7}{c}{CCSD} \\
D & 10.76 & 14.36 & 25.69 & 13.33 & 33.58 & 9.15 & 9.69 \\
T & 4.08 & 1.38 & 5.82 & 2.92 & 9.43 & 3.07 & 3.25 \\
Q & 1.80 & 0.15 & 1.23 & 1.17 & 2.85 & 1.24 & 1.31 \\
5 & 0.30 & 0.11 & 0.56 & 0.60 & 0.56 & 0.66 & 0.68 \\
  & \multicolumn{7}{c}{CCSD-F12 (AKW)}  \\
D & 5.11 & 1.15 & 13.22 & 3.22 & 17.49 & 3.09 & 3.97 \\
T & 0.85 & 0.44 & 3.21 & 0.52 & 3.81 & 0.62 & 0.73 \\
Q & 0.41 & 0.15 & 0.75 & 0.71 & 0.95 & 0.62 & 0.77 \\
5 & 0.19 & 0.11 & 0.33 & 0.27 & 0.48 & 0.32 & 0.31 \\ 
  & \multicolumn{7}{c}{CCSD-F12 (PV1)} \\
D & 5.32 & 1.50 & 13.22 & 3.22 & 17.65 & 3.33 & 2.28 \\
T & 0.82 & 0.49 & 3.21 & 0.52 & 2.98 & 0.57 & 0.83 \\
Q & 0.36 & 0.16 & 0.75 & 0.71 & 0.67 & 0.62 & 0.77 \\
5 & 0.14 & 0.11 & 0.33 & 0.27 & 0.43 & 0.32 & 0.31 \\ \hline \hline
\end{tabular}
\caption{Mean unsigned BSIEs (kJ/mol) of the reaction energies of the HJO12 benchmark set.}
\label{tbl:hjo-componentwise-error}
$^a$Includes CABS singles correction in the case of CCSD-F12.
\end{table}

\begin{table}[]
\resizebox{\textwidth}{!}{
\begin{tabular}{c|rrr|rrr|rrr|rrr}
\hline \hline

 & \multicolumn{3}{c|}{D} & \multicolumn{3}{c|}{T} & \multicolumn{3}{c|}{Q} & \multicolumn{3}{c}{5} \\
 & $X$Z & a$X$Z & $X$Z/a$X$Z & $X$Z & a$X$Z & $X$Z/a$X$Z & $X$Z & a$X$Z & $X$Z/a$X$Z & $X$Z & a$X$Z & $X$Z/a$X$Z \\ \hline
 & \multicolumn{12}{c}{CCSD}  \\
$\langle \delta \rangle$ & 30.80 & 1.01 & -0.68 & 8.79 & 1.38 & 0.92 & 2.41 & -0.07 & -0.25 & 0.28 & -0.05 & -0.09 \\
$\langle|\delta|\rangle$ & 33.58 & 9.15 & 9.69 & 9.43 & 3.07 & 3.25 & 2.85 & 1.24 & 1.31 & 0.56 & 0.66 & 0.68 \\
$\sigma$ & 25.39 & 13.17 & 13.83 & 7.15 & 4.44 & 4.57 & 2.20 & 1.62 & 1.72 & 0.64 & 0.81 & 0.83 \\
max $|\delta|$ & 60.83 & 30.35 & 30.84 & 18.31 & 11.10 & 11.30 & 4.96 & 3.79 & 3.85 & 1.53 & 1.80 & 1.82 \\
 & \multicolumn{12}{c}{CCSD-F12 (AKW)}  \\
$\langle \delta \rangle$ & 14.02 & 2.24 & 3.08 & 2.84 & -0.24 & -0.68 & 0.35 & -0.62 & -0.77 & 0.08 & -0.22 & -0.26 \\
$\langle|\delta|\rangle$ & 17.49 & 3.09 & 3.97 & 3.81 & 0.62 & 0.73 & 0.95 & 0.62 & 0.77 & 0.48 & 0.32 & 0.31 \\
$\sigma$ & 16.44 & 2.94 & 3.72 & 3.80 & 0.73 & 0.59 & 1.16 & 0.36 & 0.41 & 0.63 & 0.30 & 0.27 \\
max $|\delta|$ & 37.36 & 5.93 & 8.41 & 9.45 & 1.19 & 1.73 & 2.72 & 1.40 & 1.61 & 1.52 & 0.59 & 0.63 \\
 & \multicolumn{12}{c}{CCSD-F12 (PV1)}  \\
$\langle \delta \rangle$ & 14.35 & 2.54 & 1.71 & 2.06 & -0.29 & -0.72 & 0.07 & -0.62 & -0.77 & 0.03 & -0.22 & -0.26 \\
$\langle|\delta|\rangle$ & 17.65 & 3.33 & 2.28 & 2.98 & 0.57 & 0.83 & 0.67 & 0.62 & 0.77 & 0.43 & 0.32 & 0.31 \\
$\sigma$ & 16.50 & 3.30 & 2.34 & 3.04 & 0.64 & 0.68 & 0.86 & 0.36 & 0.43 & 0.57 & 0.30 & 0.27 \\
max $|\delta|$ & 35.12 & 7.30 & 5.34 & 6.88 & 1.20 & 1.72 & 1.71 & 1.40 & 1.62 & 1.33 & 0.59 & 0.62 \\ \hline \hline
\end{tabular}
}
\caption{Statistical measures of BSIEs (kJ/mol) of single- and dual-basis CCSD and CCSD-F12 reaction energies of the HJO12 benchmark set.$^a$}
$^a$ $\langle\delta\rangle$,  $\langle|\delta|\rangle$, $\sigma$, and max $|\delta|$ denote mean signed, mean unsigned, standard deviation, and maximum unsigned values of BSIE, $\delta \equiv E - E_\text{CBS}$.
\label{tbl:hjo12-data}
\end{table}

\cref{tbl:hjo-componentwise-error} shows the mean unsigned BSIE of CCSD and CCSD-F12 reaction energies and their component breakdowns for the small-molecule set of 12 reactions from Ref. \citenum{VRG:helgaker:2000:}; for reference, the HJO12 set and the geometries are listed in the Supporting Information. 
Detailed statistics of total CCSD and CCSD-F12 BSIEs are listed in \cref{tbl:hjo12-data}.

{\bf CCSD}. Unlike the absolute energies, where the addition of diffuse AOs led to incremental reductions of the CCSD BSIEs (on the order of $20\%$), here the impact of diffuse AOs is dramatic for the component (HF \emph{and} correlation) and total CCSD reaction energies;
e.g., with $X=\{2,3\}$ the mean unsigned BSIE of the total CCSD reaction energies is reduced by a factor of more than 3. These reductions are less pronounced for the HF and correlation energy components, but nevertheless are still substantial. Even for the quadruple-zeta basis the inclusion of the diffuse AOs reduced the mean unsigned BSIE of total CCSD reaction energies by a factor of more than 2. Only for the quintuple-zeta basis does the impact of diffuse AOs become negligible. The impact of diffuse AOs on mean {\em signed} BSIE is even more striking (\cref{tbl:hjo12-data}): reductions by a factor of 6\textendash30 were observed. Equally striking is that the mean signed BSIEs are so much smaller than the unsigned counterparts when diffuse AOs are present in the basis; the respective values with the a$\{\text{D},\text{T},\text{Q},5\}$Z OBS are \{1.01,1.38,-0.07,-0.05\} kJ/mol and  \{9.15, 3.07, 1.24, 0.66\} kJ/mol. Such dichotomy is not observed for the $X$Z basis sets. The diffuse AOs seem to help ``center'' the distribution of the reaction BSIEs near zero, but their effect on the ``width'' of the distribution, as characterized by the standard deviation $\sigma$, is less than a factor of 2 (and rapidly decreasing with $X$) for the conventional CCSD.

{\bf CCSD-F12}. Just like for the absolute energies, the impact of diffuse AOs is much more pronounced for CC-F12 reaction energies than for the standard non-F12 counterparts. Regardless of which CABS singles correction is used, the $X$Z$\to$a$X$Z OBS transition reduces mean unsigned BSIE by a factor of more than 5 for $X=2,3$; for larger $X$ the improvements are modest.
Unlike CCSD, where the addition of diffuse AOs reduced the standard deviation of BSIEs by a factor of at most 2, the standard deviations of CCSD-F12 BSIEs reduce by a factor of 2\textendash5 in the  $X$Z$\to$a$X$Z OBS transition. The conventional wisdom that the BSIEs of triple-zeta F12 energies are comparable to the standard quintuple-zeta counterpart\cite{VRG:tew:2007:PCCP,VRG:valeev:2008:JCP,VRG:adler:2007:JCP,VRG:zhang:2012:JCTC} holds here also, as long as the OBS includes diffuse AOs:
the respective \{aTZ CCSD-F12, a5Z CCSD\} mean unsigned BSIEs are \{0.57, 0.66\} kJ/mol. Without diffuse AOs the improvement from the F12 terms in only equivalent to 1 cardinal number of OBS: the respective \{TZ CCSD-F12, QZ CCSD\} mean unsigned BSIEs are \{2.98, 2.85\} kJ/mol.

{\bf Dual-Basis CCSD and CCSD-F12}.
Just like the absolute electronic energies, the total reaction energies obtained with the $X$Z/a$X$Z DB approach closely track the single-basis a$X$Z results, with the differences between the respective mean unsigned BSIEs smaller than the mean unsigned BSIEs themselves. The differences between the two variants of the CABS singles correction, AKW and PV1, are almost entirely negligible for triple- and higher-zeta basis sets (as expected) and relatively small for the single-basis double-zeta BSIEs. The use of PV1 correction in DB-CCSD-F12, however, produces noticeably smaller BSIE with the double-zeta basis; namely, the mean unsigned BSIE of 2.28 kJ/mol is obtained with the PV1 approach, compared to 3.97 kJ/mol for the AKW counterpart. This is not unexpected, since the BSIE of the HF energy and the magnitude of the conventional singles amplitudes in the dual-basis approach are the largest for the double-zeta basis, hence the benefit of the PV1 over the AKW correction is expected to be most noticeable. Note that the component breakdown of dual-basis BSIEs is meaningless (see \cref{sec:technical}). 

\subsection{Noncovalent Interaction Energies}\label{sec:results:s66}

Lastly we examined the standard S66 benchmark of small-molecule noncovalent interaction energetics\cite{VRG:rezac:2011:JCTCa,VRG:brauer:2016:PCCP,VRG:kesharwani:2018:AJC}.
\Cref{tbl:s66-combined} displays statistical measures of the BSIEs of the double- and triple-zeta CCSD and CCSD-F12 interaction energies; for definitions of HF and CCSD CBS estimates see \cref{sec:technical}.

{\bf CCSD}. The inclusion of diffuse AOs does not lower the BSIE of total CCSD binding energies with either double- or triple-zeta OBS. The analysis of component BSIEs reveals why. Basis augmentation decreases the BSIE of HF binding energies by a factor of \{2,4\} for \{DZ,TZ\} OBS, but the augmentation changes the sign of correlation BSIE and increases its magnitude in both cases, which results in an overall increase of all statistical measures of CCSD BSIE. Without explicit correlation or extrapolation the CCSD BSIEs are simply too large to be useful, whether the diffuse AOs are included or not.

\begin{table}[htbp]
\centering
\begin{tabular}{c|rrrrrrr}
\hline\hline 
 & \multicolumn{2}{c|}{HF$^b$} & \multicolumn{2}{c|}{correlation} & \multicolumn{3}{c}{total} \\ 
 & \multicolumn{1}{|c}{$X$Z} & \multicolumn{1}{c|}{a$X$Z} & $X$Z & \multicolumn{1}{c|}{a$X$Z} & $X$Z & a$X$Z & $X$Z/a$X$Z \\ \hline
 & \multicolumn{7}{c}{$X=\text{D}$} \\ \hline
& \multicolumn{7}{c}{CCSD} \\
$\langle \delta \rangle$ & -1.509 & -0.680 & 0.622 & -1.040 & -0.887 & -1.719 & -1.650 \\
$\langle|\delta|\rangle$ & 1.509 & 0.680 & 0.625 & 1.043 & 0.973 & 1.719 & 1.652 \\
$\sigma$ & 0.809 & 0.360 & 0.337 & 0.803 & 0.882 & 1.130 & 1.168 \\
max $|\delta|$ & 3.096 & 1.610 & 1.357 & 3.091 & 2.597 & 4.675 & 4.590 \\
& \multicolumn{7}{c}{CCSD-F12 (AKW)} \\
$\langle \delta \rangle$ & -1.044 & -0.441 & 0.030 & -0.300 & -1.014 & -0.742 & -0.708 \\
$\langle|\delta|\rangle$ & 1.044 & 0.441 & 0.258 & 0.301 & 1.014 & 0.742 & 0.708 \\
$\sigma$ & 0.520 & 0.226 & 0.302 & 0.153 & 0.753 & 0.353 & 0.390 \\
max $|\delta|$ & 2.035 & 1.016 & 0.603 & 0.700 & 2.638 & 1.551 & 1.510 \\
& \multicolumn{7}{c}{CCSD-F12 (PV0)} \\
$\langle \delta \rangle$ & -0.679 & -0.032 & 0.030 & -0.300 & -0.649 & -0.333 & -0.256 \\
$\langle|\delta|\rangle$ & 0.679 & 0.094 & 0.258 & 0.301 & 0.652 & 0.333 & 0.270 \\
$\sigma$ & 0.298 & 0.112 & 0.302 & 0.153 & 0.516 & 0.197 & 0.242 \\
max $|\delta|$ & 1.247 & 0.311 & 0.603 & 0.700 & 1.726 & 1.011 & 1.048 \\ 
&  \multicolumn{7}{c}{CCSD-F12 (PV1)} \\
$\langle \delta \rangle$ & -0.648 & 0.028 & 0.030 & -0.300 & -0.618 & -0.273 & -0.182 \\
$\langle|\delta|\rangle$ & 0.648 & 0.122 & 0.258 & 0.301 & 0.620 & 0.277 & 0.220 \\
$\sigma$ & 0.260 & 0.145 & 0.302 & 0.153 & 0.444 & 0.202 & 0.254 \\
max $|\delta|$ & 1.335 & 0.376 & 0.603 & 0.700 & 1.596 & 0.976 & 1.013 \\ \hline
 & \multicolumn{7}{c}{$X=\text{T}$} \\ \hline
& \multicolumn{7}{c}{CCSD} \\
$\langle \delta \rangle$ & -0.554 & -0.147 & 0.047 & -0.617 & -0.507 & -0.764 & -0.738 \\
$\langle|\delta|\rangle$ & 0.554 & 0.147 & 0.096 & 0.617 & 0.513 & 0.764 & 0.738 \\
$\sigma$ & 0.271 & 0.082 & 0.105 & 0.331 & 0.340 & 0.399 & 0.411 \\
max $|\delta|$ & 1.109 & 0.450 & 0.278 & 1.540 & 1.157 & 1.990 & 1.960 \\
& \multicolumn{7}{c}{CCSD-F12 (PV0)} \\
$\langle \delta \rangle$ & -0.194 & -0.046 & -0.091 & -0.195 & -0.285 & -0.241 & -0.225 \\
$\langle|\delta|\rangle$ & 0.194 & 0.046 & 0.125 & 0.195 & 0.286 & 0.241 & 0.225 \\
$\sigma$ & 0.078 & 0.036 & 0.138 & 0.085 & 0.170 & 0.111 & 0.122 \\
max $|\delta|$ & 0.380 & 0.204 & 0.398 & 0.406 & 0.624 & 0.610 & 0.578 \\
& \multicolumn{7}{c}{CCSD-F12 (PV1)} \\
$\langle \delta \rangle$ & -0.105 & -0.030 & -0.091 & -0.195 & -0.196 & -0.225 & -0.211 \\
$\langle|\delta|\rangle$ & 0.120 & 0.033 & 0.126 & 0.195 & 0.198 & 0.225 & 0.211 \\
$\sigma$ & 0.104 & 0.031 & 0.138 & 0.085 & 0.106 & 0.105 & 0.116 \\
max $|\delta|$ & 0.342 & 0.162 & 0.398 & 0.408 & 0.494 & 0.570 & 0.528 \\
\hline\hline
\end{tabular}
\\
$^a$ $\langle\delta\rangle$,  $\langle|\delta|\rangle$, $\sigma$, and max $|\delta|$ denote mean signed, mean unsigned, standard deviation, and maximum unsigned values of BSIE, $\delta \equiv E - E_\text{CBS}$.\\
$^b$Includes CABS singles correction in the case of CCSD-F12. 
\caption{Statistical measures of BSIEs (kcal/mol) of single- and dual-basis CCSD and CCSD-F12 reaction energies of the S66 benchmark set.$^a$} 
\label{tbl:s66-combined} 
\end{table}

{\bf CCSD-F12: Double-zeta OBS}. The impact of diffuse AOs on CCSD-F12 binding energies is even more nuanced. A clear net BSIE reduction from the basis augmentation was observed only for the double-zeta basis: the mean unsigned BSIE of total CCSD-F12 binding energies obtained with \{AKW, PV0, PV1\} CABS singles corrections reduced from \{1.014, 0.652, 0.620\} to \{0.742, 0.333, 0.277\} kcal/mol, respectively, i.e. by a factor of \{ 1.37, 1.96, 2.34\}. The significant reduction of the BSIE due to the use of the more elaborate PV CABS singles corrections indicates that the HF BSIE is a significant contributor to the overall error. The massive combined impact of the diffuse AOs \emph{together with} CABS singles correction  can be gleaned by comparing the HF \{DZ, aDZ\} BSIEs to their HF+CABS counterparts: \{1.509, 0.680\} kcal/mol mean unsigned HF BSIE are reduced to \{1.044, 0.441\}, \{0.679, 0.094\}, and \{0.648 0.122\} kcal/mol when corrected by AKW, PV0, and PV1 CABS singles corrections, respectively. It is this major reduction of the HF basis set error due to the presence of diffuse AOs together with the use of PV CABS singles corrections that allows for the remarkably small overall CCSD-F12(PV) BSIEs with the aDZ basis! Another important observation is the misleadingly-weak impact of diffuse AOs on the mean unsigned BSIE of the correlation contribution; the corresponding mean signed BSIE changes significantly, by nearly 0.3 kcal/mol. However, diffuse AOs significantly reduce the standard deviation of the correlation BSIE by a factor of $\sim 2$ (from 0.302 to 0.153 kcal/mol).

{\bf CCSD-F12: Triple-zeta OBS}. For the triple-zeta basis no net improvement from the basis augmentation is observed for the mean unsigned BSIE. Although the HF BSIEs are still significantly reduced by the CABS singles correction combined with the basis augmentation, the total BSIE is controlled by its correlation contribution. The diffuse AOs do not significantly change the mean BSIE. but this effect is a combination of a reduction of the HF BSIE and an increase of the correlation BSIE. Just as observed for the double-zeta basis, the augmentation increases mean signed and unsigned correlation BSIE but it significantly reduced its standard deviation. Lastly, while the best mean unsigned CCSD-F12 BSIE obtained with aDZ and aTZ basis are comparable (0.277 and 0.225 kcal/mol, respectively), the corresponding standard and maximum deviations are significantly smaller for the triple-zeta OBS, 0.105 and 0.570 kcal/mol, vs. 0.202 and 0.976 kcal/mol for the double-zeta counterpart.

{\bf CABS Singles: PV vs AKW}.
Recall that for the HJO12 reaction energy benchmark we did not observe a significant improvement from the use of the PV1 correction, so it is likely that the performance gains from the PV corrections are specific to the noncovalent interaction energetics. Nevertheless, it is encouraging that the more elaborate PV corrections result in better numerical performance. The perceptive reader might have realized the potential issue of comparing the PV-corrected HF energies to the AKW-corrected and CBS HF estimates. Indeed, the PV corrections are defined using CCSD singles amplitudes, which should not enter the model for correcting the HF BSIE. Thus, comparisons of CABS-corrected HF energies should be viewed as shedding light on the anatomy of the BSIE of the total CCSD-F12 energy, rather than as a means of obtaining better estimates of the HF CBS limit. This also indicates that the CBS limit of the total CC energy is best reached directly by correcting the basis set incompleteness of the full CC Lagrangian, rather than by trying to obtain CBS limits of HF and correlation energies separately.

{\bf Dual-basis CCSD and CCSD-F12}. Similarly to absolute electronic energies and HJO12 reaction energies, dual-basis $X$Z/a$X$Z formulations of CCSD and CCSD-F12 closely approach the single-basis a$X$Z reference results. The difference between them is always negligible in relation to the overall BSIE. This suggests that dual-basis methods can be used effectively for studies of noncovalent interactions in large systems, where the direct use of diffuse AOs in mean-field orbital optimization may be too poorly conditioned. 

\section{Summary}\label{sec:summary}

In this work we examined many reasons why diffuse AOs are important for accurate computation of energies even of neutral systems with wave function methods. The diffuse AOs are important for reaching the basis set limit of the mean-field (Hartree-Fock) energies, especially when combined with the perturbative (CABS) corrections for the basis set incompleteness. While for standard (non-F12) models of correlation the impact of diffuse AOs on correlation energies is relatively small compared to the impact of incrementing the cardinal (zeta) number of the basis, for explicitly-correlated (F12) models the former is much closer to the latter due to much smaller overall BSIE in the F12 methods. With the optimal choice of geminal lengthscales the impact of diffuse AOs on the CCSD-F12 correlation energies is a remarkably large percentage ($>50\%$) of the corresponding CCSD value, weakly dependent on $X$ and the atomic number (\cref{fig:abs_e:ratio-diffuse-moloptbeta}). This suggests that the correlation physics for which diffuse AOs are important is largely complementary to the short-range correlation represented by the F12 geminals. Recently revised CC-F12 geminal lengthscales\cite{VRG:powell:2025:JCTC} help maintain this ``separation of concerns'' (\cref{fig:abs_e:ratio-diffuse-optbeta}) using fixed $X$-specific lengthscales that are better than the original MP2-F12-optimized geminal parameters.\cite{VRG:peterson:2008:JCP}

Analysis of the impact of diffuse AOs was aided by the novel dual-basis CC-F12 method which enables the use of different AO bases for describing the reference (occupied) and correlating (virtual/unoccupied) orbitals. The use of a non-augmented cc-pV$X$Z basis for the former and augmented aug-cc-pV$X$Z basis for the latter allowed quantification of the impact of the diffuse AOs on the conventional and F12 components of the correlation energy. To account for the significant magnitude of conventional 1-body (singles) amplitudes in such a framework, it was necessary to reformulate the so-called CABS singles correction that accounts for the basis set incompleteness of the reference energy. The new CABS singles correction (PV1) turned out to be useful not only in the context of dual-basis CCSD-F12 but also in the context of standard CCSD-F12 applications, particularly for describing the energetics of noncovalent interactions.
The dual-basis CC-F12 formalism allowed us to compute highly accurate absolute and relative energies with diffuse bases and explicit correlation while circumventing the use of diffuse AOs in the course of the SCF orbital optimization, which can be challenging for extended systems due to the high condition number of the basis. We conjecture that dual-basis CC-F12 methods can be broadly useful for approaching the CBS limit in extended systems and in solids robustly (along the line of similar recent work\cite{VRG:maschio:2020:JCTC,VRG:maschio:2024:JPCA}).

\begin{suppinfo}

HJO12 benchmark set (reactions, molecular geometries).

\end{suppinfo}

\section*{Funding}
This work was supported by the U.S. Department of Energy via award DE-SC0022327. The work of SRP was partially supported by the Institute for Critical Technologies and Applied Science at Virginia Tech.

This work is dedicated to late Prof. John F. Stanton, a humble, curious, insightful, and inspiring giant of our field whose impact will live on forever.

\begin{appendices}

\section{Notation and Standard Definitions}
\label{sec:notation}
In this work, we largely follow a covariant tensor notation of many-body quantum chemistry \cite{VRG:harris:1981:PRA,VRG:kutzelnigg:1982:JCP}
. Matrix elements of one-, two- and higher-body operators are denoted as follows:
\begin{align}
\bra{p} \hat{o}(1) \ket{q} \equiv & \int \phi_p^*(1) \hat{o}(1) \phi_q(1) d1 \equiv o^q_p \\
\bra{p_1 p_2} \hat{o}(1,2) \ket{q_1 q_2} \equiv & \int \phi_{p_1}^*(1) \phi_{p_2}^*(2) \hat{o}(1,2) \phi_{q_1}(1) \phi_{q_2}(2) d1 d2 \equiv o^{q_1 q_2}_{p_1 p_2}, \quad \text{etc.}
\end{align}
The bra/ket indices are thus typeset in subscript/superscript positions. The Einstein summation convention is implied, unless sums over any index are shown explicitly: namely, summation is implied over every symbol that appears once in subscript and once in superscript position a given {\em product} of tensor elements.

Tensor notation is also used for strings of fermionic creation and annihilation operators. That is, particle-conserving 1- and 2-body operators are denoted as
\begin{align}
    a^p_q \equiv & a^\dagger_p a_q \\
    a^{pq}_{rs} \equiv & a^\dagger_p a^\dagger_q a_s a_r,
\end{align}
where $a^\dagger_p/a_p$ creates/annihilates a particle in $p$, respectively.

Following the convention, we used $m, n, \dots$, $a, b, \dots$, $p, q, \dots$, and $i, j, \dots$,  for the (spin-)orbitals in the reference (occupied) set, correlating (virtual/unoccupied) set, their union, and active occupied set, respectively; $\alpha, \beta, \dots$ for virtual orbitals in the complete basis; and $\alpha', \beta', \dots$ for virtual orbitals in the complete basis that do not belong to the conventional $\{a\}$ set.

The matrix elements of the 1- and 2-particle parts of the Hamiltonian
are denoted by $h^p_q$ and $g^{pq}_{rs}$, respectively,
\begin{align}
    F^p_q \equiv \, h^p_q + \left(g^{pr}_{qs} - g^{pr}_{sq} \right) \gamma^s_r,
\end{align}
where $\gamma^s_r$ denotes elements of 1-RDM, are the familiar matrix elements of the Fock operator.
The bars over 2-particle integrals denote antisymmetrization:
\begin{align}
\bar{g}^{pq}_{rs} \equiv & g^{pq}_{rs} - g^{pq}_{sr}.
\end{align}

\section{Spin-adapted equations of the PV CABS Singles Corrections}
\label{sec:pvcabseqs}

The open-shell amplitude equations are 

\begin{subequations}
\begin{align}
\frac{\partial \mathcal{L}}{\partial \lambda^{\alpha'}_m} \overset{\mathrm{PV0}}{=} & \bra{0} a^m_{\alpha'} \left(\hat{F} + [\hat{H}, \hat{T}_1] + [\hat{F}, \hat{T}_1'] \right) \ket{0} \nonumber \\
\frac{\partial \mathcal{L}}{\partial \lambda^{\alpha'}_M} = & F^M_{\alpha'} + \delta^M_K F^C_{\alpha'} t^K_C + \bar{g}^{MC}_{\alpha'K} t^K_C + g^{M \bar{C}}_{\alpha' \bar{K}} t^{\bar{K}}_{\bar{C}} + F^{\beta'}_{\alpha'} t^M_{\beta'} - F^M_{N} t^N_{\alpha'}  ,\\
\frac{\partial \mathcal{L}}{\partial \lambda^{\bar{\alpha'}}_{\bar{M}}} = & F^{\bar{M}}_{\bar{\alpha'}} + \delta^{\bar{M}}_{\bar{K}} F^{\bar{C}}_{\bar{\alpha'}} t^{\bar{K}}_{\bar{C}} + \bar{g}^{\bar{M} \bar{C}}_{\bar{\alpha'} \bar{K}} t^{\bar{K}}_{\bar{C}} + g^{\bar{M} C}_{\bar{\alpha'} K} t^{K}_{C} + F^{\bar{\beta'}}_{\bar{\alpha'}} t^{\bar{M}}_{\bar{\beta'}} - F^{\bar{M}}_{\bar{N}} t^{\bar{N}}_{\bar{\alpha'}}  ,
\label{eq:pv0:os}
\\
\frac{\partial \mathcal{L}}{\partial \lambda^{\alpha}_m} \overset{\mathrm{PV1}}{=} & \bra{0} a^m_{\alpha} \left(\hat{Q}'\left(\hat{F} + [\hat{H}, \hat{T}_1]\right) + [\hat{F}, \hat{T}_1'] \right) \ket{0} \nonumber \\
\frac{\partial \mathcal{L}}{\partial \lambda^{\alpha}_M} = & \delta^{\alpha'}_\alpha \left(F^M_{\alpha'} + F^C_{\alpha'} t^K_C \delta^M_K + \bar{g}^{MC}_{\alpha' K} t^K_C + \bar{g}^{M\bar{C}}_{\alpha' \bar{K}} t^{\bar{K}}_{\bar{C}} + \right) + F^{\beta}_{\alpha} t^M_{\beta} - F^M_{N} t^N_{\alpha}, \\
\frac{\partial \mathcal{L}}{\partial \lambda^{\bar{\alpha}}_{\bar{M}}} = & \delta^{\bar{\alpha'}}_{\bar{\alpha}} \left(F^{\bar{M}}_{\bar{\alpha'}} + F^{\bar{C}}_{\bar{\alpha'}} t^{\bar{K}}_{\bar{C}} \delta^{\bar{M}}_{\bar{K}} + \bar{g}^{\bar{M} \bar{C}}_{\bar{\alpha'} \bar{K}} t^{\bar{K}}_{\bar{C}} + g^{\bar{M} C}_{\bar{\alpha'} K} t^{K}_C + \right) + F^{\bar{\beta}}_{\bar{\alpha}} t^{\bar{M}}_{\bar{\beta}} - F^{\bar{M}}_{\bar{N}} t^{\bar{N}}_{\bar{\alpha}}
\label{eq:pv1:os}
\end{align}
\label{eq:pv:os}
\end{subequations}

and the open-shell energy expressions are

\begin{subequations}
\begin{align}
    E^{(2)}_{\hat{T}_1'} \overset{\mathrm{PV0}}{=} & \bra{0} \hat{T}_1^{\prime \dagger} \left(\hat{F} + [\hat{H}, \hat{T}_1] \right) + \hat{T}_1^\dagger [\hat{H},\hat{T}_1'] \ket{0} \nonumber \\
    = & t_M^{\alpha'} F_{\alpha'}^M + \Re t_M^{\alpha'} F^C_{\alpha'} t^K_C \delta^M_K + \Re t_M^{\alpha'} \bar{g}^{MC}_{\alpha'K} t^K_C + \Re t_{\bar{M}}^{\bar{\alpha'}} g^{\bar{M}C}_{\bar{\alpha'}K} t^K_C \\
    + & t_{\bar{M}}^{\bar{\alpha'}} F_{\bar{\alpha'}}^{\bar{M}} + \Re t_{\bar{M}}^{\bar{\alpha'}} F^{\bar{C}}_{\bar{\alpha'}} t^{\bar{K}}_{\bar{C}} \delta^{\bar{M}}_{\bar{K}} + \Re t_{\bar{M}}^{\bar{\alpha'}} \bar{g}^{\bar{M}\bar{C}}_{\bar{\alpha'}\bar{K}} t^{\bar{K}}_{\bar{C}} + \Re t_{M}^{\alpha'} g^{M \bar{C}}_{\alpha'\bar{K}} t^{\bar{K}}_{\bar{C}}  \label{eq:e_pv0:os} \\
    E^{(2)}_{\hat{T}_1'} \overset{\mathrm{PV1}}{=} & \bra{0} \hat{T}_1^{\prime \dagger} \hat{Q}' \left(\hat{F} + [\hat{H}, \hat{T}_1] \right)  + \hat{T}_1^\dagger [\hat{H}, \hat{Q}' \hat{T}_1'] \ket{0} \nonumber \\
    = & t_M^{\alpha'} F_{\alpha'}^M + \Re t_M^{\alpha'} F^C_{\alpha'} t^K_C \delta^M_K + \Re t_M^{\alpha'} \bar{g}^{MC}_{\alpha' K} t^K_C + \Re t_{\bar{M}}^{\bar{\alpha'}} g^{\bar{M}C}_{\bar{\alpha'} K} t^K_C \\
    + & t_{\bar{M}}^{\bar{\alpha'}} F_{\bar{\alpha'}}^{\bar{M}} + \Re t_{\bar{M}}^{\bar{\alpha'}} F^{\bar{C}}_{\bar{\alpha'}} t^{\bar{K}}_{\bar{C}} \delta^{\bar{M}}_{\bar{K}} + \Re t_{\bar{M}}^{\bar{\alpha'}} \bar{g}^{\bar{M} \bar{C}}_{\bar{\alpha'} \bar{K}} t^{\bar{K}}_{\bar{C}} + \Re t_{M}^{\alpha'} g^{M\bar{C}}_{\alpha' \bar{K}} t^{\bar{K}}_{\bar{C}} \label{eq:e_pv1:os}
\end{align}
\label{eq:e_pv:os}
\end{subequations}

where the indices here are spin indices and those without bars ($\alpha, M, C,$ etc.) represent alpha orbitals and those with bars ($\bar{\alpha}, \bar{M}, \bar{C},$ etc.) represent beta orbitals. 

\end{appendices}

\bibliography{vrgrefs, srprefs}

\begin{tocentry}

\includegraphics[]{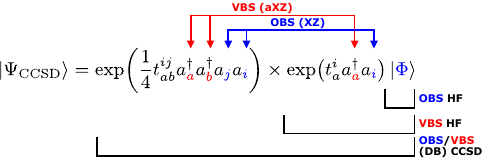} 

\end{tocentry}

\end{document}


\maketitle

\section{Molecular Geometries}
Below are molecular geometries for the 16 molecules included in the HJO12 set of reaction energies. 

\begin{verbatim}
**** C2H2 ****
4

C    0.00000000  0.00000000  0.00000000
H    0.53082419 -0.91982046 -0.00007633
C   -0.60180068  1.04280963  0.00008654
H   -1.13262593  1.96262948  0.00015646
**** C2H4 ****
6

C    0.00000000  0.00000000  0.00000000
H   -0.51681895 -0.94945204 -0.00000231
C   -0.66527966  1.15280700  0.00009567
H   -1.74592825  1.18036780  0.00026157
H   -0.14846071  2.10225903  0.00019522
H    1.08064858 -0.02756080 -0.00019680
**** CH2 ****
3

C    0.00000000  0.00000000  0.00000000
H    1.10154209  0.10979084  0.00000000
H   -0.33641512  1.05464395  0.00000000
**** CH2O ****
4

C    0.00000000  0.00000000  0.00000000
O    1.20399999  0.00016351  0.00000000
H   -0.58030540  0.93565306  0.00000000
H   -0.58005125 -0.93581064  0.00000000
**** CH4 ****
5

C    0.00000000  0.00000000  0.00000000
H    0.00000000  0.00000000  1.08600000
H    1.02389074  0.00000000 -0.36199966
H   -0.51194552 -0.88671526 -0.36199977
H   -0.51194552  0.88671526 -0.36199977
**** CO ****
2

C 0 0 0
O 0 0 1.129
**** CO2 ****
3

C 0 0 0
O 0 0 -1.16
O 0 0 1.16
**** F2 ****
2

F 0 0 0
F 0 0 1.411
**** H2 ****
2

H 0 0 0
H 0 0 0.742
**** H2O ****
3

O    0.00000000  0.00000000  0.00000000
H    0.95522327  0.05828807  0.00000000
H   -0.29083042  0.91173827  0.00000000
**** HF **** 
2

H         0.000000    0.000000   -0.455093
F         0.000000    0.000000    0.455093
**** HNC ****
3

N 0 0 0
H 0 0 -0.995
C 0 0 1.1693
**** HCN ****
3

C 0 0 0
H 0 0 -1.066
N 0 0 1.154
**** N2 ****
2

N 0 0 0
N 0 0 1.098
**** NH3 ****
4

N    0.00000000  0.00000000  0.00000000
H    0.25959754 -0.89817105  0.38473222
H    0.64770131  0.67357746  0.38587228
H   -0.90779370  0.22388130  0.38538931
**** HNO ****
3

N    0.00000000  0.00000000  0.00000000
H    0.51921299 -0.89970198  0.16630762
O   -0.27596158  0.47818028  1.07445466
\end{verbatim}

\section{HJO12 Set of Reaction Energies}

\begin{table}[htb]
\centering
    \begin{tabular}{rr@{}l|rr@{}l} \hline \hline
        no. & \multicolumn{2}{r|}{reaction} & no. & \multicolumn{2}{r}{reaction} \\ \hline
         1 & \ce{CO + H2} $\rightarrow$ & \xspace \ce{CH2O} & 7 & \ce{HCN + 3H2} $\rightarrow$& \xspace \ce{CH4 + NH3} \\
         2 & \ce{N2 + 3H2} $\rightarrow$ & \xspace \ce{2NH3} & 8 & \ce{HNO + 2H2} $\rightarrow$& \xspace \ce{NH3 + H2O} \\
         3 & \ce{C2H2 + H2} $\rightarrow$ & \xspace \ce{C2H4} & 9 & \ce{C2H2 + 3H2} $\rightarrow$& \xspace \ce{CH4 + NH3}  \\
         4 & \ce{CO2 + 4H2} $\rightarrow$ & \xspace \ce{CH4 + 2H2O} & 10 & \ce{CH2 + H2} $\rightarrow$& \xspace \ce{CH4}  \\
         5 & \ce{CH2O + 2H2} $\rightarrow$ & \xspace \ce{CH4 + H2O} & 11 & \ce{F2 + H2} $\rightarrow$& \xspace \ce{2HF} \\
         6 & \ce{CO + 3H2} $\rightarrow$ & \xspace \ce{CH4 + H2O} & 12 & \ce{2CH2} $\rightarrow$& \xspace \ce{C2H4} \\ \hline \hline
    \end{tabular}
    \caption{Reactions in the HJO12 reaction set. See Ref. \citenum{VRG:helgaker:2000:} for further details.}
    \label{tbl:hjo12}
\end{table}